\documentclass[pra,twocolumn,nofootinbib,a4paper]{revtex4-1}
\usepackage{graphicx}
\usepackage{enumitem}
\usepackage{amsmath}
\newcommand{\ket}[1]{|{#1}\rangle}
\newcommand{\bra}[1]{\langle{#1}|}

\def\Xint#1{\mathchoice
   {\XXint\displaystyle\textstyle{#1}}%
   {\XXint\textstyle\scriptstyle{#1}}%
   {\XXint\scriptstyle\scriptscriptstyle{#1}}%
   {\XXint\scriptscriptstyle\scriptscriptstyle{#1}}%
   \!\int}
\def\XXint#1#2#3{{\setbox0=\hbox{$#1{#2#3}{\int}$}
     \vcenter{\hbox{$#2#3$}}\kern-.5\wd0}}

\def\dashint{\Xint-}
\usepackage{amsfonts}
\begin{document}

\title{Bose-Fermi dualities for arbitrary one-dimensional quantum systems in the universal low energy regime}

\author{Manuel Valiente}
\affiliation{Institute for Advanced Study, Tsinghua University, Beijing 100084, China}

\begin{abstract}
I consider general interacting systems of quantum particles in one spatial dimension. These consist of bosons or fermions, which can have any number of components, arbitrary spin or a combination thereof, featuring low-energy two- and multiparticle interactions. The single-particle dispersion can be Galilean (non-relativistic), relativistic, or have any other form that may be relevant for the continuum limit of lattice theories. Using an algebra of generalized functions, statistical transmutation operators that are genuinely unitary are obtained, putting bosons and fermions in a one-to-one correspondence without the need for a short-distance hard core. In the non-relativistic case, low-energy interactions for bosons yield, order by order, fermionic dual interactions that correspond to the standard low-energy expansion for fermions. In this way, interacting fermions and bosons are fully equivalent to each other at low energies. While the Bose-Fermi mappings do not depend on microscopic details, the resulting statistical interactions heavily depend on the kinetic energy structure of the respective Hamiltonians. These statistical interactions are obtained explicitly for a variety of models, and regularized and renormalized in the momentum representation, which allows for theoretically and computationally feasible implementations of the dual theories. The mapping is rewritten as a gauge interaction, and one-dimensional anyons are also considered.
\end{abstract}
\pacs{}

\maketitle

\section{Introduction}
Ensembles of interacting quantum particles exhibit a remarkable variety of fascinating phenomena. These occur at various scales, ranging from the microscopic few-body regime to the macroscopic world of materials. Interesting few-body physics can be found in systems as varied as atoms and molecules \cite{Blume2012}, small nuclei \cite{Bedaque2002,Frederico} and even in the description of solids, with excitons \cite{Wang2018} and trions \cite{Tempelaar2019}, together with magnons \cite{Zapf2014}, being prime examples of the latter. Of particular interest is the Efimov effect \cite{Efimov1970}, consisting of the existence of a series of three-body bound states that obey a discrete scaling law when three bosons interact in such a way that the two-body scattering length diverges, i.e. in the unitary limit. These were sought for intensively in nuclear physics for many years \cite{Fedorov1994}, and were first observed with ultracold Cs by Kraemer {\it et al.} \cite{Kraemer2006}. They have since been confirmed in atomic $^4\mathrm{He}$ as well \cite{Kunitski2015}, and have been predicted to occur for three magnons in a solid \cite{Nishida2013}. The Efimov effect represents a paradigmatic example of universality \cite{Braaten2006}, in which microscopic details of the multiparticle interactions are irrelevant -- Cs-Cs interactions are of a completely different nature from magnon-magnon interactions -- and low-energy scattering properties are sufficient to describe the phenomenon \cite{Frederico,Greene2017}. At the few-body level, universality of interactions is directly linked to the fact that, as far as two- and multi-particle scattering amplitudes are concerned, short distance details of the interactions, and therefore the wave functions, are not important, since the amplitudes are only related to the physics at long distances. These interactions, which still describe microscopic phenomenology, are the subject of study of low-energy effective field theory (EFT) of interactions \cite{Bedaque2002,BraatenEFT,ValienteZinnerEFT}. In many-body physics, universality shows in a number of ways. The connection with few-body universality is given by the use of effective low-energy interactions to simplify their {\it a priori} extremely complicated description. For instance, while in a fully microscopic description, $^{87}\mathrm{Rb}$ atoms interact with each other via detailed, strong Born-Oppenheimer potentials, it is possible to describe a dilute, ultracold gas of $^{87}\mathrm{Rb}$ atoms using only the two-body scattering length \cite{Bloch2008}, which produces an effective zero-range interaction given by the Fermi-Huang-Yang pseudopotential \cite{Huang1987}. Mean-field theory, as well as Bogoliubov theory, which would miserably fail using the original interaction, has immense predictive power in the physics of Bose-Einstein condensates \cite{Pethick}, be it $^{87}\mathrm{Rb}$ or any other many-boson system, provided it remains dilute. Remarkably, both weak and strong, non-perturbative interactions, in the sense of EFTs, can often be effectively described by the same universal low-energy macroscopic theory in the many-body case. A prominent example of this type of universality includes Luttinger liquid theory \cite{Haldane1981}. In one spatial dimension, gapless many-body systems are described, at low energies, by Luttinger liquid theory. This applies to both bosons and fermions, with or without spin, and (essentially) arbitrarily weak or strong microscopic interactions \cite{Giamarchi}. The ``trick'' in these effective macroscopic theories -- which do not solve the microscopic many-body problem -- is that, while their non-universal parameters require solving the microscopic many-body problem fully non-perturbatively, once these are fixed, it is possible to extract large amounts of information about these systems.

Quantum systems in one spatial dimension are currently attracting great interest, especially in the context of ultracold atoms \cite{Cazalilla2011,Sowinski2019}, which can be effectively constrained to move in one spatial dimension by means of anisotropic trapping. Although of importance in other fields, systems of ultracold atoms in (quasi-) one dimension (i) can be composed of spinless, spinful or multicomponent bosons and fermions \cite{Haller2009,Paredes2004,Serwane2011,Wenz2013,Murmann2015,Hulet2020}, or mixtures thereof \cite{Sowinski2019}; (ii) can be trapped in a variety of geometries \cite{Greiner2017,BlochSSH}; (iii) can consist of as few as two or three particles \cite{Greiner2017,Serwane2011,Reynolds2020} as well as many \cite{Kinoshita2004}; and (iv) their effective low-energy interactions can be tuned at will by means of Feshbach \cite{Hulet2020,Chin2010,Zhai2015} and confinement-induced resonances \cite{Olshanii1998,Haller2010,Sala2012}.

A particularly interesting feature of interacting one-dimensional quantum systems is the thin line separating the properties of bosons and fermions, especially in the spinless case. Non-relativistic spinless bosons and fermions whose two-body interactions feature a short-distance hard core, otherwise arbitrary, are equivalent, or dual to each other, as shown by Girardeau 60 years ago \cite{Girardeau}. His result is known as the Bose-Fermi mapping theorem for hard core particles. If the hard core has zero range and there are no other interactions, bosons are dual to non-interacting fermions and the system is exactly solvable, and is in the so-called Tonks-Girardeau limit, first realized experimentally in Ref.~\cite{Paredes2004}. It was not until 1999 that a duality relation between non-relativistic soft core spinless bosons and fermions was found by Cheon and Shigehara \cite{CheonShigehara}. In particular, $N$ spinless bosons with zero-range two-body Dirac delta interactions (corresponding to lowest order EFT for bosons) of arbitrary strength, known as the Lieb-Liniger model \cite{LiebLiniger}, are dual to $N$ spinless fermions with lowest order interactions in the odd-wave channel. The Cheon-Shigehara mapping was used to find a duality relation between non-relativistic spin-$1/2$ fermions with lowest order even- and odd-wave pseudopotentials and two-component bosons by Girardeau and Olshanii \cite{Girardeau2004}, while a point hard core Bose-Fermi mapping was found for non-relativistic spin-$1$ bosons \cite{Deuretzbacher2008}. The derivation of the duality relation for soft core spinless bosons and fermions, Ref.~\cite{CheonShigehara}, relies heavily on microscopic details, in particular short range boundary conditions. There, a clever guess of regularized Dirac delta interactions and their zero-range limit, in the position representation, are crucial. The resulting interaction for fermions is then given by a zero-range pseudopotential that is chosen to match the dual bosonic scattering phase shifts. In Ref.~\cite{Girardeau2004}, the Cheon-Shigehara mapping, together with symmetry arguments, was sufficient to map non-relativistic spin-$1/2$ fermions and two-component bosons with the same lowest-order interactions. If one wishes to generalize soft core Bose-Fermi mappings in one dimension to arbitrary dispersion (beyond non-relativistic), spin or internal structure and arbitrary low energy interactions, within the EFT paradigm, it is clear that the approach of Refs.~\cite{Girardeau,CheonShigehara,Girardeau2004}, where pseudopotentials are guessed rather than derived, quickly becomes impossible to handle. This is because (i) the short-range boundary conditions imposed by a particular low energy interaction depend heavily on the single-particle dispersion; (ii) boundary conditions are increasingly complicated for higher order interactions that may include effective range or three-particle effects; (iii) the approach relies in one way or another in the analytical knowledge of exact wave functions, which are certainly out of reach for complicated non-integrable systems with more than two particles. Moreover, with the exception of the Lieb-Liniger model, the resulting pseudopotentials appear difficult to use in either approximate or exact numerical treatments, due to their highly singular nature.

Technically, what hinders the ability to produce a fully universal Bose-Fermi mapping in one dimension is the fact that the signum function, $S(x)=1$ for $x>0$ and $-1$ for $x<0$, is a distribution or generalized function \cite{Schwartz}; this distribution is present in all one-dimensional Bose-Fermi mappings \cite{Girardeau,CheonShigehara,Girardeau2004,Deuretzbacher2008}. As such, its value at the origin is undefined, which can be problematic unless the wave functions vanish at the coalescence point of two particles (hard core condition \cite{Girardeau}). In order for a duality relation to be valid, the two dual systems should be in one-to-one correspondence, i.e. there must be a unitary transformation relating them. Specifically, the signum function must be elevated to a unitary operation, that is, one must be able to set $[S(x)]^2=1$ for all values of $x$ including the origin, regardless of whether $S(x)$ itself is defined or not at $x=0$, and this must be done in a mathematically consistent manner.

In this article, I find the most general one-to-one mapping between bosons and fermions in one spatial dimension with arbitrary low energy interactions, not restricted to pairwise forces. The mapping applies to any internal structure (single- or multicomponent) and spin. It is also valid for arbitrary single-particle dispersion, including non-relativistic, relativistic and continuum limits of lattice Hamiltonians. This is done by regarding the unavoidable generalized functions that appear in the unitary transformations and multiparticle interactions as members of an algebra of distributions constructed by Shirokov in Ref.~\cite{Shirokov}. The goal of the algebra is to provide a direct regularization and renormalization procedure in the position representation, and to provide a mathematically rigorous framework to elevate the map between bosons and fermions to a unitary operator. Since the algebra yields formal expressions that are not of much use unless exact solutions are provided, all interactions are given in the momentum representation and regularized according to standard cutoff schemes. These provide a theoretically and computationally simple prescription for the practical use of the duality relations. At low energies, it is shown here that non-relativistic soft core bosons and fermions are equivalent to each other order by order in their respective low-energy EFTs, explicitly up to lowest-order three-body interactions. This is to say that the Bose-Fermi mapping does not take the dual system out of the low-energy scattering regime. As a corollary, it is possible to describe soft core bosons, fermions and hard core bosons (which are equivalent to fermions due to Girardeau's mapping \cite{Girardeau}), at low energies, using either effective representation: near the Tonks-Girardeau limit, one can use the fermionic EFT perturbatively, while in the opposite limit, it is most convenient to use the bosonic representation. Since, as the order of the EFT description increases, the interactions eventually become too singular, separable, regular terms may be necessary, and the duality relations also apply to these. The unitary Bose-Fermi mapping operator is also given explicitly for general multicomponent or spinful systems. Two examples of duality relations are given, namely the continuum limit of the fermionic Su-Schrieffer-Heeger (SSH) model \cite{SSHpaper1,SSHpaper2} near half-filling, which features first derivatives only at the single-particle level, and non-relativistic spin-$1/2$ fermions described by Yang's model \cite{Yang1967}. The statistical interaction in the continuum SSH model is shown to differ significantly from the non-relativistic case, and is shown to be renormalizable in its momentum representation. The duality relation from fermions to two-component bosons is therefore satisfied, a fact that is also shown explicitly. For Yang's model, I show how grid computations can be easily performed by discretizing the dual two-component bosonic Hamiltonian on a lattice, where three-body calculations in the continuum limit give identical results for both dual representations. The general Bose-Fermi mapping, being unitary, is finally written as a gauge interaction, and one-dimensional anyons are briefly considered.

This article is an extended and augmented version of an accompanying Letter \cite{Letter}, and is organized as follows. In Sect.~\ref{SectionShirokov}, I introduce Shirokov's algebra of generalized functions. In Sect.~\ref{SectionSTO}, I define Bose-Fermi mappings, which I call statistical transmutation operators, with desirable properties, and explicitly find them; the concept of statistical interaction is also defined and the formal duality relations are obtained. In Sect.~\ref{SectionEFT}, low-energy two- and three-body interactions for non-relativistic bosons and fermions are considered, and some renormalizability issues for fermions are pointed out. Duality relations between non-relativistic spinless bosons and fermions at low energies are studied in detail in Sect.~\ref{SectionDuality1}. Multicomponent and spinful systems are considered in Sect.~\ref{SectionDuality2}. In Sect.~\ref{SectionGauge}, the duality transformations are written as a gauge interaction, and one-dimensional anyons are considered. Conclusions and some important consequences are presented in Sect.~\ref{SectionConclusions}. Finally, some further technical details are given in four Appendices.

\section{Algebra of distributions}\label{SectionShirokov}
The transformations that map bosonic wave functions onto fermionic ones and viceversa are generally singular \cite{Girardeau}. In particular, they involve mathematical distributions (or generalized functions) in the sense of Schwartz \cite{Schwartz}, that is, the usual, linear theory of distributions. As will be seen below, the action of the transformed, or dual Hamiltonian of either the bosonic or fermionic representations, involves products of distributions. These products, unfortunately, do not in general represent a mathematical distribution in the sense of Schwartz \cite{Schwartz2}. To overcome this issue, a nonlinear theory of distributions is required. Nonlinear theories of distributions \cite{Shirokov,Colombeau,Egorov} are concerned with the construction of associative algebras which, from a practical point of view, regularize and renormalize expressions such as $[\delta(x)]^2$, $S(x)\delta(x)$ or $\delta(x)\delta'(x)$, where $\delta(x)$ and $S(x)$ are, respectively, the Dirac delta and signum distributions. The reason behind constructing an algebra $\mathcal{A}$ corresponds to the reasonable requirement that if two distributions belong to $\mathcal{A}$, so does their product. Associativity allows to define the algebraic product operation pairwise. The field over which the algebra is defined is obviously the complex numbers $\mathbb{C}$, and $\mathcal{A}$ is constructed such that differentiation respects Leibniz's rule and that complex conjugation (adjoint) works in the usual way. Although Colombeau's algebra \cite{Colombeau} is the best known in the mathematical literature, especially because it is commutative, Shirokov's algebra $\mathcal{U}$ \cite{Shirokov}, which is much simpler but non-Abelian, was constructed with quantum mechanics in mind. Therefore, from here on, every distribution encountered is treated as a member of $\mathcal{U}$. The defining properties of Shirokov's algebra are the following:
\begin{align}
  \delta^{(m)}(x)\delta^{(n)}(x)&=0 \hspace{0.2cm} \forall m,n\ge 0,\label{deltasquare}\\
  \{S(x),\delta(x)\}&=0,\label{anticommutator}\\
         [S(x)]^2&=1 \hspace{0.2cm} \forall x,\label{Ssquare}
\end{align}
where $\delta^{(n)}(x)=\partial^n_x\delta(x)$ and $\{\cdot,\cdot\}$ is the anticommutator. Note that Eq.~(\ref{anticommutator}) follows from Eq.~(\ref{Ssquare}), since $\partial_x [S(x)S(x)]=0=\{S(x),\delta(x)\}$.

As a simple example of how the properties of $\mathcal{U}$ work, take the following bosonic two-body wave function in the relative coordinate $x$, $\psi^{k}_{\mathrm{B}}(x)=\sin(k|x|+\theta_k)$, and its fermionic dual $\psi^{k}_{\mathrm{F}}(x)=S(x)\psi^{k}_{\mathrm{B}}(x)$. The usual assertion that $|\psi^{k}_{\mathrm{B}}(x)|^2=|\psi^{k}_{\mathrm{F}}(x)|^2$ $\forall x$ is only correct after the algebra is chosen: property (\ref{Ssquare}) is not at all trivial, and $[S(x)]^2$ is otherwise undefined at $x=0$. 

\section{Statistical transmutation operators}\label{SectionSTO}
In this section I introduce the concept of statistical transmutation operators (STO), which formally transform bosonic functions into fermionic functions and viceversa. For a particular one-dimensional $N$-body system, I require the STO to be (i) linear; (ii) unitary; (iii) energy independent; and (iv) local \footnote{Recall that an operator $\hat{O}$ is local iff $\bra{x'}\hat{O}\ket{x} \propto \delta(x-x')$.}.

With the above conditions, I define a boson-to-fermion STO $\mathcal{T}$ such that any fermionic state $\ket{\chi}$ becomes bosonic after application of $\mathcal{T}$. It then holds that $\ket{\chi}=\mathcal{T}^{\dagger}\ket{\psi}$, and $\ket{\psi}=\mathcal{T}\ket{\chi}$. If $\ket{\psi}$ is an eigenstate of $H_{\mathrm{B}}$ with eigenenergy $E$, then $\ket{\chi}$ is an eigenstate of $H_{\mathrm{F}}=\mathcal{T}H_{\mathrm{B}}\mathcal{T}^{\dagger}$ with the same energy. 

In the single-channel two-body example of the previous section, we have
\begin{equation}
  \bra{x'}\mathcal{T}x\rangle = S(x)\delta(x-x')=\bra{x'}\mathcal{T}^{\dagger}x\rangle,
\end{equation}
that is, the STO is local. Moreover, it is unitary, since $\bra{x'}\mathcal{T}\mathcal{T}^{\dagger}x\rangle = \bra{x'}\mathcal{T}^{\dagger}\mathcal{T}x\rangle = \delta(x-x')$. Note, again, that for unitarity to hold the use of the algebra is required, in particular Eq.~(\ref{Ssquare}).  

It is also necessary to establish whether a particular one-dimensional quantum system admits an STO at all. Without loss of generality, I will use an $N$-boson system as reference, whose dynamics is described by Hamiltonian $H_{\mathrm{B}}=H_0+V_{\mathrm{B}}$, where $H_0$ is a general single-particle operator including all kinetic energy terms, which can be multichannel and contain single-particle external potentials, and $V_{\mathrm{B}}$ is the interaction, not necessarily pairwise, but possibly multichannel in accordance with the structure of $H_0$. It can be shown that every quantum one-dimensional system, whether single- or multi-channel in nature, admits an STO. To see this, denote by $\mathbf{\xi}_i=(x_i,\mathbf{m}_i)$ ($i=1,\ldots,N$) the degrees of freedom of particle $i$, with $x_i$ the position and $\mathbf{m}_i$ a vector containing the internal degrees of freedom in the multichannel case. An $N$-body wave function $\psi$ satisfying bosonic statistics can be unitarily transformed into a wave function $\chi$ satisfying fermionic statistics by means of the following local STO $\mathcal{T}$
\begin{align}
  \bra{\mathbf{\xi}_1',\ldots,\mathbf{\xi}_N'}\mathcal{T}\ket{\mathbf{\xi}_1,\ldots,\mathbf{\xi}_N}&=\delta(\mathbf{x}-\mathbf{x}')S_N(\mathbf{x})\prod_{i=1}^N\delta_{\mathbf{m}_i,\mathbf{m}_i'},\label{diagonalSTO}\\
  S_N(\mathbf{x})\equiv \prod_{i<j=1}^N S(x_i-x_j).
\end{align}
The unitarity of operator $\mathcal{T}$, which is diagonal in all degrees of freedom, is guaranteed by the fact that, within Shirokov's algebra, $[S(x)]^2=1$ $\forall x$. The simplicity of the diagonal STO of Eq.~(\ref{diagonalSTO}) makes it a very appealing choice. However, except for the single-channel case, in which the STO (\ref{diagonalSTO}) is actually unique -- a consequence of the locality condition -- multichannel systems admit more than one STOs. These are however related by (symmetric) unitary transformations.

Duality relations are useful if it is possible to obtain strongly-coupled bosonic (fermionic) solutions from a weakly-coupled fermionic (bosonic) theory. Therefore, it is convenient to cast the transformed Hamiltonian, again without loss of generality using bosons as a reference, in the usual form of kinetic + potential terms. The original Hamiltonian $H_{\mathrm{B}}$ for the bosonic system is given by
\begin{equation}
  H_{\mathrm{B}}=H_0+V_{\mathrm{B}},
\end{equation}
where, again, all kinetic energy operators are contained in $H_0$. The transformed Hamiltonian $H_F=\mathcal{T}H_{\mathrm{B}}\mathcal{T}^{\dagger}$ can then be rewritten as
\begin{equation}
H_{\mathrm{F}}\equiv H_0+W_{\mathrm{F}}+V_{\mathrm{F}},\label{HF1}
\end{equation}
where $W_{\mathrm{F}}$ is the totally antisymmetric projection of $W$, with
\begin{align}
W&=[\mathcal{T},H_0]\mathcal{T}^{\dagger},\label{statisticalinteraction}\\
V_{\mathrm{F}}&=\mathcal{T}V_{\mathrm{B}}\mathcal{T}^{\dagger}.\label{ordinaryinteraction}
\end{align}
I will call $V_{\mathrm{F}}$ the ordinary dual fermionic interaction, and $W$ the statistical interaction.

\section{Non-relativistic single-channel effective field theories}\label{SectionEFT}
Before reviewing existing duality relations and establishing new ones, it is important to understand how bosons and fermions interact at low energies. This is especially relevant because, at least in the single-channel case, it would be very convenient from a theoretical point of view if low-energy bosonic interactions map, via the STO, order-by-order onto the respective low-energy fermionic interactions. If this is the case, as I will show in Sect.~\ref{SectionDuality1}, then it is possible to assert that at low energies fermions and bosons are completely equivalent and, therefore, the description of universal low-energy physics in one dimension can be done from either fermionic or bosonic side, whichever is most convenient for the particular application. For instance, $^4\mathrm{He}$ atoms tightly confined to (quasi-) one dimension, have a strongly repulsive (infinite) two-body core at short distances \cite{Bertaina,DelMaestro} but the two-body scattering length is very large and positive \cite{ValienteOhberg}. The existence of a short-distance hard core makes this system of $^4\mathrm{He}$ atoms be equivalent to fermions with the same Hamiltonian \cite{Girardeau}, suggesting that the low-energy physics at low densities should be described by a fermionic effective field theory (EFT). However, the large scattering length means that the fermionic EFT is in the strong-coupling (attractive) regime \cite{CheonShigehara,ValienteZinnerEFT}, making the description of the system highly non-perturbative. The use of the more convenient soft-core bosonic EFT, which is weakly-coupled for large scattering lenghts, including three-body forces that stabilize its liquid phase \cite{ValienteOhberg}, can only be theoretically justified if the fermionic and bosonic representations of the low-energy EFT are equivalent. 

In the following subsections, I consider low-energy interactions in the two- and three-body sectors for bosons and fermions, with Hamiltonian
\begin{equation}
  H=\sum_{i}\frac{p_i^2}{2m}+\sum_{i<j}V^{(2)}_{ij}+\sum_{i<j<l}V^{(3)}_{ijl}+\ldots,
\end{equation}
where $V^{(n)}_{ij\ldots}$ represents an $n$-body interaction. I restrict these interactions to the smooth, hyperspherically symmetric case, that is, their momentum representations $V^{(2)}(q)$, with $q$ the relative momentum exchange, only depend on $q$ through $q^2$, while $V^{(3)}$ only depends on the three-body hypermomentum $q_H$ via $q_H^2$ and not on the three-body angular variables.

\subsection{Bosonic low-energy interactions}
In the non-relativistic single-channel case, developing low-energy EFTs simply consists, at the bare level, of expanding the interactions in power series of the momentum transfer. The two-body interaction $V^{(2)}_{ij}(q)$ is expanded as
\begin{equation}
  V^{(2)}(q)=v_0+v_2q^2+O(q^4).\label{V2q}
\end{equation}
Since the momentum representation of the ``true'' smooth two-body interaction is given by
\begin{equation}
  \bra{k'}V^{(2)}\ket{k}=V^{(2)}(k-k')\equiv V^{(2)}(q),
\end{equation}
its even-wave part $V^{(2)}_{\mathrm{e}}(k',k)$, which has a non-vanishing effect on bosonic wave functions, takes the form
\begin{equation}
  V^{(2)}_{\mathrm{e}}(k',k)=v_0+v_2(k^2+k'^2)+O(q^4).\label{BosonTwoBody}
\end{equation}
If the interaction above is described to leading order (LO) only ($v_{n>0}\equiv 0$), then $v_0$ must be replaced by its renormalized value $g_0$, which is obtained by fixing the scattering length $a$ to the actual scattering length obtained using the ``true'' interaction $V^{(2)}$, as $g_0=-2\hbar^2/ma$ \cite{LiebLiniger}. An $N$-boson system with LO interactions corresponds to the Lieb-Liniger model \cite{LiebLiniger}. If the next-to-leading order (NLO) term in Eq.~(\ref{BosonTwoBody}) is also retained, then $v_0$ cannot be set to $g_0$ as above as the NLO term produces momentum-independent terms in the scattering amplitude \cite{ValienteZinnerEFT}. 

The three-body interaction is expanded in the same way as the two-body potential, i.e.
\begin{equation}
  V^{(3)}(q_{\mathrm{H}})=w_0+w_2q_{\mathrm{H}}^2+O(q_{\mathrm{H}}^4).\label{V3q}
\end{equation}
For a three-particle system, $q_{\mathrm{H}}^2=\sum_{i<j=1}^3(k_{ij}-k_{ij}')^2$, where $k_{ij}$ is the relative momentum for the pair $(i,j)$. The bosonic interaction becomes, from Eq.~(\ref{V3q}),
\begin{equation}
  V_{\mathrm{B}}^{(3)}(\mathbf{k}',\mathbf{k})=w_0+w_2\sum_{i<j=1}^3(k_{ij}^2+k_{ij}'^2)+\ldots,\label{BosonThreeBody}
\end{equation}
where $\mathbf{k}$ is shorthand for $(k_{12},k_{23},k_{13})$. Also note that, in Eqs.~(\ref{V3q}) and (\ref{BosonThreeBody}), total momentum ($K=k_1+k_2+k_3$) conservation is implicitly assumed, i.e. the full three-body interaction is $2\pi \delta(K-K')V^{(3)}(q_{\mathrm{H}})$.

The few-body phenomenology associated with the effective interactions above is as follows. As I already mentioned, LO two-body interactions correspond to a zero-range potential that describes the scattering length only. The LO+NLO interaction can describe the even-wave scattering phase shifts up to the effective range \cite{ValienteZinnerEFT}. Unfortunately, the term $\propto (k^2+k'^2)$ introduces, already at the two-body level, severe complications and restrictions in its renormalization procedure, in one dimension \cite{ValienteZinnerEFT}, just as in higher dimensions \cite{Cohen,BraatenEFT}. Since the NLO term represents an off-shell interaction (it depends on momentum), it can be neglected at the two-body level and the effective range can be included by allowing energy-dependence in the LO coupling constant $g_0=g_0(E)$ \cite{vanKolck,Blume}. Off-shell contributions of the two-body NLO term for three and higher particle numbers, are typically included perturbatively \cite{vanKolck,Bazak}, or else a regularised separable version, such as $\propto (k^2+k'^2)\exp[-k^2/\Lambda^2-k'^2/\Lambda^2]$, with fixed $\Lambda$, may be employed non-perturbatively. Within a Lagrangian formalism, using field redefinitions \cite{WeinbergBook}, it is possible to trade, for more than two particles, the NLO off-shell potential for a LO three-body contact force \cite{SchwenkRMP}, bypassing this issue in a different way. The LO three-body interaction, with bare coupling constant $g_0^{(3)}$ replacing $w_0$ in Eq.~(\ref{BosonThreeBody}), has been shown to be important for realistic two-body interactions in one dimension with large scattering length \cite{ValienteThreeBody}, where the three-body interaction and its three-body range correction dominate low-energy physics \cite{ValienteThreeBody,ValientePastukhov,Guijarro,Drut,Maki,Nishida,Pricoupenko,Drut2}.

The LO three-body interaction gives rise to a single logarithmic ultraviolet (UV) divergence when calculating the scattering amplitude \cite{ValienteThreeBody,Guijarro,Drut}, which is easily renormalized away in favor of the three-body scattering length $a_3$ \cite{Pricoupenko} or, equivalently, a three-body momentum scale $Q_*$ ($\propto 1/a_3$) \cite{ValienteThreeBody} beyond which the EFT description breaks down. I will use the latter convention from here on. If the three-body $T$-matrix with no two-body interactions exhibits a (Landau) pole for (repulsive) attractive interactions at $E=E_*=-\hbar^2Q_*^2/2m$, then the bare three-body coupling constant as a function of a hard hyperradial cutoff $\Lambda$ is given by \cite{ValienteThreeBody}
\begin{equation}
\frac{1}{g_0^{(3)}}=\frac{m}{\sqrt{3}\pi\hbar^2}\ln\left|\frac{Q_*}{\Lambda}\right|,
\end{equation}
and the three-body $T$-matrix at energy $z=\hbar^2k^2/2m+i0^+$ is a constant (independent of momentum) and reads \cite{ValientePastukhov}
\begin{equation}
T_3(z)=\frac{\hbar^2}{m}\frac{2\pi\sqrt{3}}{\ln\left(\frac{Q_*^2}{k^2}\right)+i\pi}.
\end{equation}

\subsection{Fermionic low-energy interactions}
Here, I consider non-relativistic spinless or spin-polarized fermions. In the two-body sector, these are affected by $V^{(2)}(q)$, Eq.~(\ref{V2q}), starting at $O(q^2)$. The first two terms (LO+NLO) of the odd-wave interaction are given by
\begin{equation}
V_{\mathrm{o}}^{(2)}(k',k)=-2v_2k'k-4v_4k'k(k^2+k'^2)+O(q^6).\label{FermionEFT1}
\end{equation}
Above, $-2v_2$ and $-4v_4$ are to be replaced by bare coupling constants $g_{1}$ and $g_{3}$, respectively.
In this case, already the LO two-body interaction gives rise to a linear UV divergence in the calculation of the amplitude \cite{ValienteZinnerEFT} and, with the appropriate relation between $g_{1}(\Lambda)$ and $g_0$, this interaction corresponds to the Cheon-Shigehara (fermions) \cite{CheonShigehara} dual to the Lieb-Liniger (bosons) \cite{LiebLiniger} interaction. The LO+NLO interaction can also be renormalized, as in the bosonic case, but suffers from the same issues and complications \cite{ValienteZinnerEFT}.

\begin{figure}[t]
\begin{center}
\includegraphics[width=0.49\textwidth]{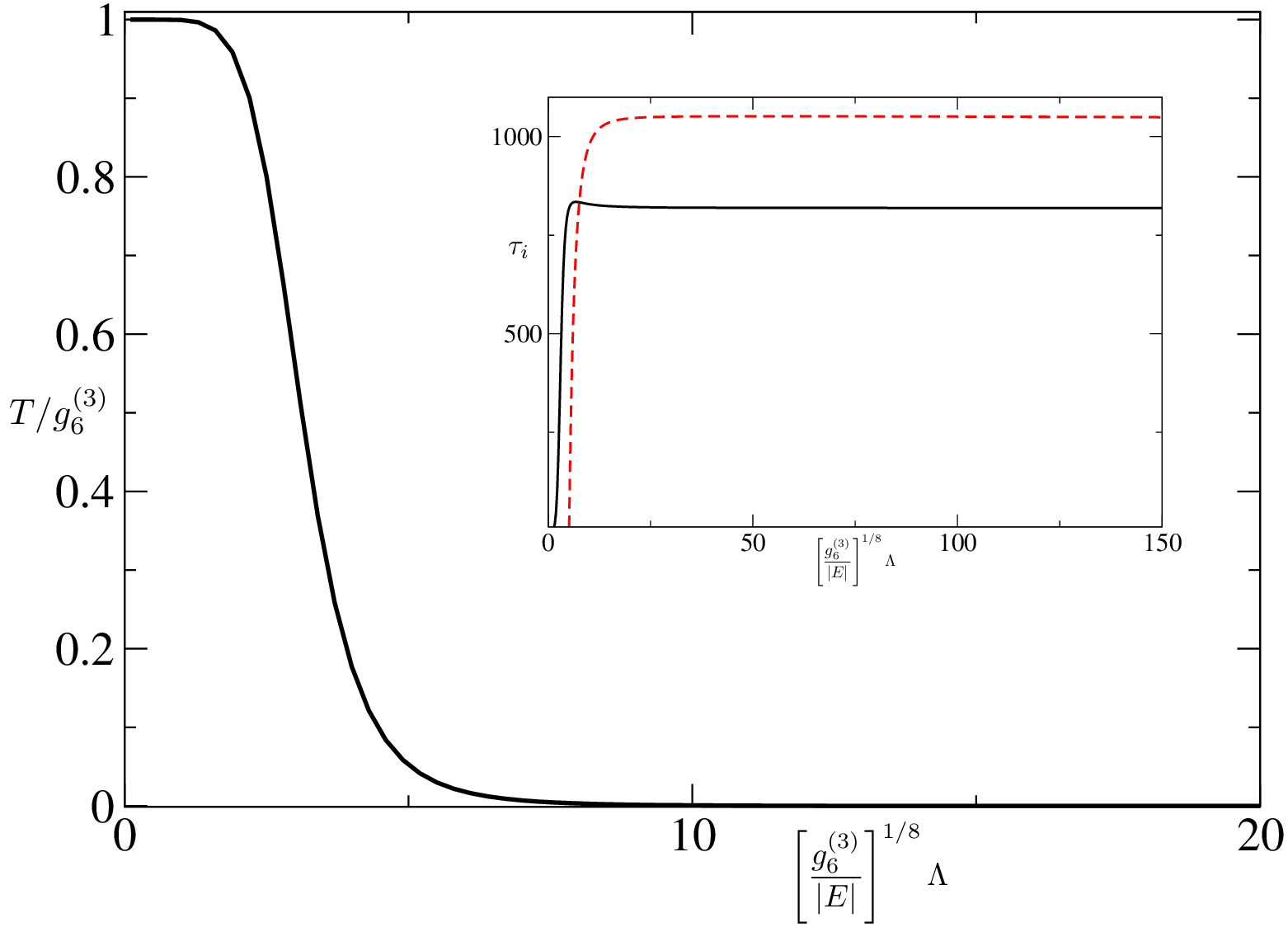}
\end{center}
\caption{Coefficient of the three-body $T$-matrix for spinless fermions with three-body interaction in Eq.~(\ref{Fermiong6}) for fixed $g_6^{(3)}$ and energy $z=E=-|E|$, as a function of the cutoff. Inset: rescaled coefficients $\tau_6$ (solid black line) and $\tau_4$ (dashed red line), see text.}
\label{fig:Tmatrix}
\end{figure}

The three-body sector is more complicated. The lowest-order three-body interaction for fermions starts only at $O(q_{\mathrm{H}}^6)$ in Eq.~(\ref{V3q}). Lower order interactions only affect bosons and distinguishable particles or mixtures. The LO three-body interaction for three fermions takes the form
\begin{equation}
  V_{\mathrm{F}}^{(3)}=g_6^{(3)}\prod_{i<j=1}^3k_{ij}k_{ij}',\label{Fermiong6}
\end{equation}
where total momentum conservation is implicitly assumed, and $g_6^{(3)}$ is the bare coupling constant. The corresponding $T$-matrix is calculated from the Lippmann-Schwinger equation $T_{3,\mathrm{F}}(z)=V_{\mathrm{F}}^{(3)}+V_{\mathrm{F}}^{(3)}G_0(z)T_{3,\mathrm{F}}(z)$, with $z$ the energy and $G_0(z)$ the non-interacting Green's function. The equation is solved in the momentum representation by $T_{3,\mathrm{F}}(z;\mathbf{k}',\mathbf{k})=T(z)\prod_{i<j=1}^3k_{ij}k_{ij}'$. Clearly, the interaction appears too singular to be renormalizable, since integrands in the calculation of the three-body $T$-matrix involve a product of $k$ from the Jacobian, $k^{-2}$ from the Green's function, and $k^6$ from the potential and $T$-matrix, for large $k$, yielding a leading UV divergence in a hard cutoff of $O(\Lambda^6)$. As I will show below, there are three subleading divergences and only one coupling constant, making this interaction non-renormalizable. To illustrate this, in Fig.~\ref{fig:Tmatrix} I plot the coefficient $T(z)$ of the $T$-matrix at total momentum $K=0$ for a fixed coupling constant $g_6^{(3)}$ and negative energy $z=E=-|E|$, as a function of the cutoff $\Lambda$. Clearly, the $T$-matrix vanishes very quickly as the cutoff is increased. In the inset, I show rescaled values of the $T$-matrix $\tau_i$, $i=6,4$, given by $\tau_6(\Lambda)=(T/g_6^{(3)})[(g_6^{(3)}/|E|)^{1/8}\Lambda]^6$ and $\tau_4=(\tau_6(\Lambda)-\tau_6(\infty))[(g_6^{(3)}/|E|)^{1/8}\Lambda]^8$, which show that the two leading UV divergences in the inverse $T$-matrix are of $O(\Lambda^6)$ and $O(\Lambda^4)$, respectively.

Analytically, this can be seen as follows. The $T$-matrix has a pole at $z=E<0$ if the following equation is satisfied
\begin{equation}
1=g_6^{(3)}\int \frac{\mathrm{d}k_1\mathrm{d}k_2}{(2\pi)^2}\frac{k_{12}^2k_{13}^2k_{23}^2}{E-\epsilon(k_1,k_2,K-k_1-k_2)},\label{fermionIE}
\end{equation}
where total momentum conservation is implicitly assumed, $k_{ij}=(k_i-k_j)/2$, and $\epsilon(k_1,k_2,k_3)=\sum_{i=1}^{3}\hbar^2k_i^2/2m$. For $E<0$, the integral in Eq.~(\ref{fermionIE}) can be bounded by changing to polar coordinates $k_1=k\cos\theta$, $k_2=k\sin\theta$, with $K=0$ without loss of generality and a hard cutoff ($k<\Lambda$), as
\begin{align}
  &\int \mathrm{d}k_1\mathrm{d}k_2 \frac{k_{12}^2k_{13}^2k_{23}^2}{|E|+\epsilon(k_1,k_2,K-k_1-k_2)}\nonumber\\
&\le \int_0^{\Lambda}\mathrm{d}kk^7(162\pi)\frac{1}{|E|+\hbar^2k^2/2m}.\label{fermionbound}
\end{align}
The divergent part of Eq.~(\ref{fermionbound}) in the UV has four terms, proportional to $\Lambda^6$, $\Lambda^4$, $\Lambda^2$ and $\ln(\Lambda/\Lambda_0)$, with $\Lambda_0$ an arbitrary, finite momentum scale to render the expression in the logarithm dimensionless.

Non-renormalizability of the three-body interaction above on its own simply means that three fermions with three-body interactions only do not interact at low energies. A duality relation between bosons and fermions in the three-body sector would show two important results: (i) the inclusion of two-body forces lifts the non-renormalizability of the fermionic LO three-body interaction and (ii) bosons in the Tonks-Girardeau (point hard core) limit are not affected by three-body interactions at low energies. Item (ii) sounds very natural but, since the bare three-body interaction is not regular, it requires a fully non-perturbative treatment. This is done in the following section.

\section{Duality in the single-channel non-relativistic case}\label{SectionDuality1}

\subsection{Statistical transmulation operator and statistical interaction}
In the structureless, single component, single channel case, the STO is unique, as outlined in Sect.~\ref{SectionSTO}. From Eq.~(\ref{diagonalSTO}), the matrix elements of the STO are given by
\begin{equation}
\bra{x_1',x_2',\ldots,x_N'}\mathcal{T}^{\dagger}x_1,x_2,\ldots,x_N\rangle = S_N(\mathbf{x})\delta(\mathbf{x}-\mathbf{x}'),\label{STO1}
\end{equation}
Using the STO in Eq.~(\ref{STO1}), I now derive the statistical interaction $W$, Eq.~(\ref{statisticalinteraction}). Its action on an arbitrary wave function, not necessarily an eigenstate of any Hamiltonian, is straightforward to obtain introducing Eq.~(\ref{STO1}) into Eq.~(\ref{statisticalinteraction}), with $H_0=\sum_ip_i^2/2m$, and we have
\begin{equation}
  W(\mathbf{x})=-\frac{\hbar^2}{2m}S_N(\mathbf{x})\left[\nabla_N^2S_N(\mathbf{x})+2\nabla_NS_N(\mathbf{x})\cdot \nabla_N\right],
\end{equation}
where $\nabla_N$ is the $N$-dimensional gradient. Explicitly, one finds
\begin{equation}
W(\mathbf{x})=-\frac{2\hbar^2}{m}\sum_{i<j=1}^NS(x_{ij})\left[\delta'(x_{ij})+2\delta(x_{ij})\partial_{x_{ij}}\right],\label{WF1channel}
\end{equation}
where $x_{ij}=x_i-x_j$. Importantly, the statistical interaction (\ref{WF1channel}) contains only pairwise interactions, which implies no further terms need to be included for more than two particles.

\subsection{Practical examples}\label{Examples}
To show how duality works from a practical operational point of view, I will begin by solving two examples, namely the two-body fermionic dual to non-interacting bosons, and the many-body bound state of fermions dual to the bosonic MacGuire state with attractive delta interactions \cite{MacGuire}.

\subsubsection{Two-body Bose-Fermi mapping}
The fermionic Hamiltonian dual to two non-interacting bosons simply reads, after separation of center of mass $X$ and relative ($x$) coordinates, using Eqs.~(\ref{HF1}) and (\ref{WF1channel}),
\begin{equation}
H_{\mathrm{F}}=-\frac{\hbar^2}{m}\partial_{x}^2-\frac{2\hbar^2}{m}S(x)\left[\delta'(x)+2\delta(x)\partial_{x}\right].\label{HF2body}
\end{equation}
The fermionic ground state (with eigenenergy $E=0$), according to the above analysis, has the form $\chi_0(x)=S(x)$. Applying $H_{\mathrm{F}}$ in Eq.~(\ref{HF2body}) on $\chi_0(x)$, I obtain
\begin{align}
  H_{\mathrm{F}}\chi_0&=-\frac{2\hbar^2}{m}\delta'(x)-\frac{2\hbar^2}{m}S(x)\left[\delta'(x)S(x)+2(\delta(x))^2\right]\nonumber\\
  &=-\frac{2\hbar^2}{m}\delta'(x)+\frac{2\hbar^2}{m}\delta'(x)=0,
\end{align}
where I have used properties (\ref{deltasquare}), (\ref{anticommutator}) and (\ref{Ssquare}) of Shirokov's algebra $\mathcal{U}$ to go from the first to the second line. Note that the statistical interaction in Eq.~(\ref{HF2body}) admits no bound states. Also, for arbitrary positive energy $E=\hbar^2k^2/m$, the scattering phase shift $\delta_{\mathrm{o}}$ ($\delta_{\mathrm{e}}$) induced for fermions (bosons) with scattering states $\chi(x)=S(x)\sin(k|x|+\delta_{\mathrm{o}})$ ($\psi(x)=\cos(k|x|+\delta_{\mathrm{e}})$) is simply $\delta_{\mathrm{o}}=\delta_{\mathrm{e}}=\pi/2$.

\subsubsection{MacGuire's state for dual fermions}
MacGuire's state for $N$ bosons is the ground state of the attractive Lieb-Liniger model, i.e.
\begin{equation}
  H_{\mathrm{B}}=\sum_{i=1}^N\frac{p_i^2}{2m}+g\sum_{i<j=1}^N\delta(x_i-x_j),\label{LiebLinigerModel}
\end{equation}
with $g<0$. The simplest way to extract the ground state of Hamiltonian (\ref{LiebLinigerModel}) is to write it in supersymmetric form \cite{DelCampo2020}. Defining $A_i=\partial_{x_i}+v_i(\mathbf{x})$, one can build a Hermitian Hamiltonian $H_{\mathrm{S}}$ as
\begin{equation}
  H_{\mathrm{S}}=\frac{\hbar^2}{2m}\sum_{i=1}^NA_i^{\dagger}A_i.
\end{equation}
Since $H_{\mathrm{S}}$ is positive semi-definite, if $A_i\psi=0$ $\forall i=1,\ldots,N$ and $\psi$ is normalizable, then $\psi$ is the ground state of $H_{\mathrm{S}}$ with energy $E=0$. For the Lieb-Liniger model, the so-called super potential $v_i$ is given by
\begin{equation}
  v_i(\mathbf{x})=\alpha\sum_{j\ne i}S(x_i-x_j),
\end{equation}
with $\alpha=-mg/2\hbar^2$. It is easy to see that the Hamiltonian $H_{\mathrm{S}}$ is given by
\begin{equation}
  H_{\mathrm{S}}=H_{\mathrm{B}}-E_0,
\end{equation}
where $H_{\mathrm{B}}$ is the Lieb-Liniger Hamiltonian (\ref{LiebLinigerModel}) and $E_0=-(mg^2/4\hbar^2)N(N+1)(N-1)/6$ is a constant to be identified with the ground state energy of $H_{\mathrm{B}}$ provided that the state $\psi$ annihilated by $A_i$ $\forall i=1,\ldots,N$ is normalizable, which is indeed the case for $g<0$. The ground state takes the well-known form \cite{MacGuire}
\begin{equation}
\psi(\mathbf{x})=\prod_{i<j=1}^N \exp(-\alpha|x_i-x_j|).\label{MacGuireGroundState}
\end{equation}
I show now explicitly that the fermionic dual to MacGuire's state (\ref{MacGuireGroundState}) is the ground state of the dual Hamiltonian $H_{\mathrm{F}}$, by writing down the transformed Hamiltonian $H_{\mathrm{S},\mathrm{F}}$ as
\begin{equation}
  H_{\mathrm{S},\mathrm{F}}=\mathcal{T}H_{\mathrm{S}}\mathcal{T}^{\dagger}=\frac{\hbar^2}{2m}\sum_i A_{i,\mathrm{F}}^{\dagger}A_{i,\mathrm{F}},
\end{equation}
with $A_{i,\mathrm{F}}=\mathcal{T}A_{i}\mathcal{T}^{\dagger}$ explicitly given by
\begin{equation}
  A_{i,\mathrm{F}}=A_i+S_N(\mathbf{x})\partial_{x_i}S_N(\mathbf{x}).
\end{equation}
The action of $A_{i,\mathrm{F}}$ on $\chi(\mathbf{x})=S_N(\mathbf{x})\psi(\mathbf{x})$, with $\psi$ in Eq.~(\ref{MacGuireGroundState}), is given by
\begin{align}
  A_{i,\mathrm{F}}\chi(\mathbf{x})&=[\partial_{x_i}S_N(\mathbf{x})]\psi(\mathbf{x})+S_N(\mathbf{x})\partial_{x_i}\psi(\mathbf{x})\nonumber\\
  &+S_N(\mathbf{x})v_i(\mathbf{x})\psi(\mathbf{x})+S_N(\mathbf{x})[\partial_{x_i}S_N(\mathbf{x})]S_N(\mathbf{x})\psi(\mathbf{x})\nonumber\\
  &=S_N(\mathbf{x})\left[\partial_{x_i}\psi(\mathbf{x})+v_i(\mathbf{x})\psi(\mathbf{x})\right]=0,
\end{align}
where I have used the anticommutativity of $S_N(\mathbf{x})$ and $\partial_{x_i}S_N(\mathbf{x})$ due to property (\ref{anticommutator}) of $\mathcal{U}$ and, in the last line, the fact that $A_i\psi=0$. Therefore, we have $H_{\mathrm{F}}\chi=E_0\chi$.

\subsection{Duality in effective field theory I:\\ Two-body sector}
I will show here how, order-by-order, bosonic low-energy EFT is equivalent, or dual to its fermionic counterpart. I begin with the bosonic two-body interaction $V_{\mathrm{e}}^{(2)}$ in Eq.~(\ref{BosonTwoBody}). In operator form, its fermionic dual $\tilde{V}_{\mathrm{o}}^{(2)}=\mathcal{T}V_{\mathrm{e}}^{(2)}\mathcal{T}^{\dagger}$, where I have used Eq.~(\ref{ordinaryinteraction}). I define fermionic two-body eigenstates $\ket{\chi_{k}}$ of $H_0$ as $\chi_k(x)=\sin(kx)$ in the relative coordinate $x=x_1-x_2$, and hardcore bosonic states $\ket{\phi_k}=\mathcal{T}\ket{\chi_k}$, with position representation $\phi_k(x)=\sin(k|x|)$. The matrix elements $\bra{\chi_{k'}}\tilde{V}_{\mathrm{o}}^{(2)}\ket{\chi_{k}}$ clearly vanish, since $\bra{\chi_{k'}}\tilde{V}_{\mathrm{o}}^{(2)}\ket{\chi_{k}}=\bra{\phi_{k'}}V_{\mathrm{e}}^{(2)}\ket{\phi_{k}}$ and, for instance to lowest order,
\begin{equation}
\bra{\phi_{k'}}V_{\mathrm{e},\mathrm{LO}}^{(2)}\ket{\phi_{k}}=g_0\int \mathrm{d}x \sin(k'|x|)\delta(x)\sin(k|x|)=0.
\end{equation}
This does not mean that the interaction potential vanishes identically (try it on interacting states of the form $S(x)\sin(k|x|+\theta_k)$, with $\theta_k$ not necessarily the correct phase shift). Instead, it means regularization is necessary when using the non-interacting fermionic basis, i.e. standard low-energy EFT. Instead of using a short-distance regularization of the Dirac delta function, e.g. $\delta_b(x)=[\delta(x-b)+\delta(x+b)]/2$, which is problematic within Shirokov's algebra \cite{Shirokov}, I work directly in the momentum representation -- which is nevertheless the goal of EFT -- and use a hard cutoff $\Lambda$ in the relative momentum. Using Eq.~(\ref{ordinaryinteraction}), and the STO, Eq.~(\ref{STO1}), which in the two-body sector is simply $S_2(x_1,x_2)=S(x_{12})$, together with the fact that one dimensional functions are split only into symmetric (S) and antisymmetric (A) parts as $f(x_{12})=f_{\mathrm{S}}(x_{12})+f_{\mathrm{A}}(x_{12})$, it is immediate to verify that
\begin{equation}
  V_{\mathrm{e}}^{(2)}(k',k)=\int \frac{\mathrm{d}q'}{2\pi}\int \frac{\mathrm{d}q}{2\pi} S_{k'-q'}^*V_{\mathrm{o}}^{(2)}(q',q)S_{k-q},\label{VBint1}
\end{equation}
where the integrals run from $-\Lambda$ to $\Lambda$ and $S_k$ is the Fourier transform of the signum function, i.e.
\begin{equation}
  S_k=\int \mathrm{d}xS(x)e^{ikx}=2i\mathrm{P.V.}\frac{1}{k},
\end{equation}
where P.V. denotes Cauchy's principal value, so that Eq.~(\ref{VBint1}) becomes
\begin{equation}
V_{\mathrm{e}}^{(2)}(k',k)=\dashint \frac{\mathrm{d}q'}{\pi}\dashint \frac{\mathrm{d}q}{\pi}\frac{V_{\mathrm{o}}^{(2)}(q',q)}{(k'-q')(k-q)},\label{VBint2}
\end{equation}
where the dashed integral sign $\dashint$ indicates it is a principal value integral. I proceed to showing how the fermionic EFT gives, order-by-order, a bosonic EFT to the same order. To LO, $V_{\mathrm{o}}^{(2)}(q',q)=\alpha q'q$ which, inserted into Eq.~(\ref{VBint2}) gives
\begin{equation}
  V_{\mathrm{e,LO}}^{(2)}(k',k)=\frac{4}{\pi^2}\alpha(\Lambda^2+O(1)).
\end{equation}
The above relation gives $\alpha=g_0\pi^2/4\Lambda^2$. In order to complete the boson-fermion correspondence, it is also necessary to extract the momentum representation of the fermionic statistical interaction $W_{\mathrm{F}}$. Using Eq.~(\ref{statisticalinteraction}) together with (\ref{STO1}), it is easy to see that
\begin{equation}
\bra{k'}W\ket{k}=\int \frac{\mathrm{d}q}{2\pi} S_{k'-q}^*\frac{\hbar^2q^2}{m}S_{k-q} - \frac{\hbar^2k^2}{m}2\pi \delta(k-k').\label{Wgen}
\end{equation}
The odd-wave projection of the statistical interaction above takes the simple form (see Appendix \ref{Appendix})
\begin{equation}
  W_{\mathrm{F}}(k',k)=-\frac{\pi\hbar^2}{m\Lambda}k'k + O(\Lambda^{-3}).\label{WFodd}
\end{equation}
Therefore, the full bare LO fermionic coupling constant that is dual to the bosonic LO interaction takes the form
\begin{equation}
g_1(\Lambda)=\frac{\pi^2g_0}{4\Lambda^2}-\frac{\pi\hbar^2}{m\Lambda}.\label{g1}
\end{equation}
To see that the two-fermion problem is renormalized in this way, notice that there is a bound state with energy $E=-mg_0^2/4\hbar^2$ for
\begin{equation}
  \frac{1}{g_1(\Lambda)}=-\int_{-\Lambda}^{\Lambda} \mathrm{d}q \frac{q^2}{|E|+\hbar^2q^2/m},
\end{equation}
which gives, as $\Lambda\to \infty$,
\begin{equation}
  \frac{1}{g_{1}(\Lambda)}=-\frac{m\Lambda}{\pi\hbar^2}-\left(\frac{m}{\hbar^2}\right)^2\frac{g_0}{4}\label{g12}
\end{equation}
Eq.~(\ref{g1}) can be inverted and expanded in powers of $\Lambda^{-1}$, obtaining Eq.~(\ref{g12}) plus terms of $O(\Lambda^{-1})$ and lower, as I wanted to show.

The order-by-order duality relation can be continued. The LO+NLO fermionic interaction has the form
\begin{equation}
  V_{\mathrm{o}}^{(2)}(k',k)=k'k\left[\alpha_2+\beta_2(k^2+k'^2)\right].\label{VF5}
\end{equation}
Introducing the above interaction in Eq.~(\ref{VBint2}), the full regularized-renormalized (in the limit $\Lambda\to\infty$) bosonic dual is given by
\begin{align}
  V_{\mathrm{e}}^{(2)}&=\frac{\alpha_2}{\pi^2}\left[4\Lambda^2+2\Lambda\ln\left|\frac{(\Lambda-k')(\Lambda-k)}{(\Lambda+k')(\Lambda+k)}\right|\right.\nonumber\\
    &\left.+\ln\left|\frac{\Lambda-k'}{\Lambda+k'}\right|\ln\left|\frac{\Lambda-k}{\Lambda+k}\right|\right]\nonumber\\
  &+\frac{\beta_2}{\pi^2}\left\{\left[2\Lambda+\ln\left|\frac{\Lambda-k'}{\Lambda+k'}\right|\right]\right.\nonumber\\
  &\left.\times\left[k^3\ln\left|\frac{\Lambda-k}{\Lambda+k}\right|+2\Lambda k^2+\frac{2\Lambda^3}{3}\right]+(k\leftrightarrow k')\right\}.\label{DualFullNLO}
\end{align}
The above interaction is dual to the odd-wave interaction (\ref{VF5}) together with the odd-wave statistical interaction (\ref{WFodd}). Eq.~(\ref{DualFullNLO}) looks complicated. However, it only contains sums of separable terms, that is, terms of the form $f_1(k')f_2(k)$, and therefore the two-body problem can be reduced to an algebraic system of equations. These are still rather cumbersome, and do not show low-energy duality in a transparent fashion. In order to see this, Eq.~(\ref{DualFullNLO}) can be expanded in powers of the momenta $k'$ and $k$, obtaining
\begin{equation}
  V_{\mathrm{e}}^{(2)}=\frac{4\alpha_2\Lambda^2}{\pi^2}+\frac{8\Lambda^4}{3\pi^2}\beta_2+\frac{8\beta_2\Lambda^2}{3\pi^2}(k^2+k'^2)+O(k^4),\label{DualBosonExpanded}
\end{equation}
where $O(k^4)$ englobes terms $\propto k'^2k^2$ and $\propto k^4$, $k'^4$.
Comparing the above with Eq.~(\ref{BosonTwoBody}), the coupling constants are na{\"i}vely related as
\begin{align}
  \beta_2&\sim\frac{3\pi^2g_2}{8\Lambda^2},\label{beta2}\\
  \alpha_2&\sim\frac{\pi^2g_0}{4\Lambda^2}-\frac{\pi^2g_2}{4},\label{alpha2}
\end{align}
where the $\sim$ symbol denotes the relation is only formal within the expansion. Unfortunately, because the above expressions come from low-energy expansions of the (already regular) effective interaction (\ref{DualFullNLO}), Eqs.~(\ref{beta2},\ref{alpha2}) would only be valid in the Born approximation, which is invalid given the fully non-perturbative nature of the problem. Duality within the EFT formalism to this order requires more microscopic input than in the LO problem studied earlier, and duality is proved fully non-perturbatively by solving the exact Lippmann-Schwinger equations for even- and odd-wave interactions in Appendix \ref{AppendixNLO}. There, it is shown that bosons with LO+NLO interactions of the form $V_e(k',k)=g_0+g_2(k'^2+k^2)$ are dual to their fermionic counterpart, of the form $V_o(k',k)+W_{\mathrm{F}}(k',k)=k'k[g_1+g_3(k'^2+k^2)]$ upon renormalization to the same scattering length and effective range. That is, if for coupling constants $g_0(\Lambda)$ and $g_2(\Lambda)$ ($\Lambda\to\infty$), one has $\psi(x)=\cos(k|x|+\delta_{\mathrm{e}})$, then it is possible to find energy-independent $g_1(\Lambda)$ and $g_3(\Lambda)$ that renormalize the two-body problem such that $\chi(x)=S(x)\psi(x)$ for all $x$. The EFT programme can be continued easily to arbitrary order but, at least in one dimension, it is rather impractical and I will not proceed any further.

It is also important to remember that, even though the main application of the duality relations is to low-energy physics, the mapping does work for regular interaction potentials. It works as well as for other types of regularizations. For instance, beyond LO, effective range effects with no energy dependent constants are easier to introduce numerically with soft, separable potentials, which are non-local. As an example, I work out an exactly solvable two-body problem with a separable potential in the even-wave channel of the type:
\begin{equation}
  \bra{x'}V_{\mathrm{e}}\ket{x}=e^{-a|x'|}V_0e^{-a|x|},
\end{equation}
with Fourier transform $V_{\mathrm{e}}(k',k)=V_0F_{\mathrm{e}}^*(k')F_{\mathrm{e}}(k)$, and
\begin{equation}
  F_{\mathrm{e}}(k)=\frac{2a}{a^2+k^2}.
\end{equation}
Its odd-wave dual is simply
\begin{equation}
  \bra{x'}V_{\mathrm{o}}\ket{x}=S(x')e^{-a|x'|}V_0e^{-a|x|}S(x),
\end{equation}
with Fourier transform $V_{\mathrm{o}}(k',k)=V_0F_{\mathrm{o}}^*(k')F_{\mathrm{o}}(k)$, and
\begin{equation}
  F_{\mathrm{o}}(k)=2\frac{k/a}{1+(k/a)^2}.
\end{equation}
The $T$-matrix with even-wave interactions takes the form
\begin{equation}
T_{\mathrm{e}}(z;k',k)=V_0t_{\mathrm{e}}(z)F_{\mathrm{e}}^*(k')F_{\mathrm{e}}(k),
\end{equation}
with
\begin{equation}
  t_{\mathrm{e}}(z)=\frac{1}{1-V_0\int\frac{\mathrm{d}q}{2\pi}\frac{\left|F_{\mathrm{e}}(q)\right|^2}{z-\hbar^2q^2/m}}.
\end{equation}
A two-boson bound state with energy $E=-|E|$ occurs when $t_{\mathrm{e}}$ has a pole, and the bound state equation reads
\begin{equation}
  \frac{\hbar^2}{mV_0}=-\frac{2a+\sqrt{m|E|/\hbar^2}}{a\sqrt{m|E|/\hbar^2}(a+\sqrt{m|E|/\hbar^2})^2}.\label{boundsep}
\end{equation}
For two fermions, the total interaction is given by $V_{\mathrm{o}}(k',k)+W_{\mathrm{F}}(k',k)$. At the bound state energy $E=-|E|$, the residue $\tau$ of the $T$-matrix can be written as
\begin{equation}
  \tau(k)=(A+BF_{\mathrm{o}}(k))k,
\end{equation}
where $A$ and $B$ are two constants. These satisfy the coupled system of equations
\begin{align}
  A&=g_{\mathrm{F}}\int\frac{\mathrm{d}q}{2\pi}\frac{q^2}{E-q^2}(A+BF_{\mathrm{o}}(q)),\\
  B&=V_0\int \frac{\mathrm{d}q}{2\pi}\frac{q^2}{E-q^2}(AF_{\mathrm{o}}(q)+B|F_{\mathrm{o}}(q)|^2),
\end{align}
where I have defined $g_{\mathrm{F}}=-\pi\hbar^2/m\Lambda$. It is tedious yet straightforward to solve the equations above, which yield, for the bound state energy, Eq.~(\ref{boundsep}). This procedure shows that, as expected, bound state problems -- typically weakly-coupled for bosons and strongly-coupled for fermions -- are easiest to handle in the bosonic representation. 

\subsection{Duality in effective field theory I.b: $s$-wave two-body collisions in three dimensions}
It is worth noting that the one-dimensional Bose-Fermi duality in the two-body sector can be used to simplify the calculation of two-body $s$-wave scattering amplitudes in three dimensions within the EFT formalism. To see this, write the $s$-wave stationary Schr{\"o}dinger equation for a three-dimensional spherically symmetric interaction $V(r)$ for the reduced radial wave function $u(r)=rR(r)$,
\begin{equation}
-\frac{\hbar^2}{m}u''(r)+V(r)u(r)=Eu(r).
\end{equation}
Since the Schr{\"o}dinger equation above is defined for $r\in[0,\infty)$, two simple extensions of $u$ to the domain $(-\infty,\infty)$ that are non-zero at the origin (i.e. solutions of zero-range EFTs) can be defined. Firstly, a continuous but non-differentiable extension that is parity symmetric, which I denote $u_{\mathrm{B}}$, which is given by $u_{\mathrm{B}}(r)=u(|r|)$ $\forall r\in (-\infty,\infty)$. And secondly, $u_{\mathrm{F}}(r)=S(r)u_{\mathrm{B}}(r)$. The scattering amplitudes for the symmetric (bosonic) and antisymmetric (fermionic) extensions, as one-dimensional problems, are exactly related order-by-order by the duality relations defined above. The $s$-wave scattering amplitude can be directly related to the one-dimensional fermionic scattering amplitude and, by duality, to the bosonic amplitude, as follows. Denote the $s$-wave projection of the central interaction $V$ by $V_s$ which, in the momentum representation, only depends on $\mathbf{k}$ and $\mathbf{k}'$ via their moduli $k$ and $k'$, respectively. The $T$-matrix $T_s$ from the $s$-wave interaction only depends on $k$ and $k'$ as well, and is a solution to the Lippmann-Schwinger equation
\begin{equation}
  T_s(z;k',k)=V_s(k',k)+\frac{1}{2\pi^2}\int_0^{\Lambda}\mathrm{d}q \frac{q^2V_s(k',q)}{z-\hbar^2q^2/m}T_s(z;q,k),
\end{equation}
where I have set a hard cutoff $\Lambda$ to regularize the integral equation. Since, moreover, the analytic continuation of $V_s$ that is continuous and differentiable satisfies $V_s(-k',k)=V_s(k',-k)=V_s(k',k)$, it is possible to extend the domain of the integration and write
\begin{equation}
  T_s(z;k',k)=V_s(k',k)+\frac{1}{4\pi^2}\int_{-\Lambda}^{\Lambda}\mathrm{d}q \frac{q^2V_s(k',q)}{z-\hbar^2q^2/m}T_s(z;q,k).\label{Ts}
\end{equation}
For one dimensional fermions, the low-energy interaction (including $W_{\mathrm{F}}$) can be written as $k'\mathcal{V}(k',k)k/2\pi$, and the $T$-matrix $T_{\mathrm{F}}(k',k)$ satisfies
\begin{equation}
  T_{\mathrm{F}}(z;k',k)=k'\frac{\mathcal{V}(k',k)}{2\pi}k+\frac{1}{4\pi^2}\int_{-\Lambda}^{\Lambda}\mathrm{d}q\frac{k'\mathcal{V}(k',q)q}{z-\hbar^2q^2/m}T_{\mathrm{F}}(z;q,k),
\end{equation}
which can be simplified by writing $T_{\mathrm{F}}(z;k',k)\equiv k'\tau(z;k',k)k$, yielding
\begin{equation}
  \tau(z;k',k)=\frac{\mathcal{V}(k',k)}{2\pi}+\frac{1}{4\pi^2}\int_{-\Lambda}^{\Lambda}\mathrm{d}q\frac{q^2\mathcal{V}(k',q)}{z-\hbar^2q^2/m}\tau(z;q,k).\label{tau}
\end{equation}
A simple comparison between Eqs.~(\ref{Ts}) and (\ref{tau}) shows that if $\mathcal{V}(k',k)=V_s(k',k)$, then
\begin{equation}
  T_s(z;k',k)=2\pi \tau(k',k).
\end{equation}
Since the $s$-wave $T$-matrix is equivalent to the one-dimensional fermionic one, it can be obtained by solving the one-dimensional bosonic problem as well (on-shell) and invoking the duality relations here derived. The bosonic representation, as already mentioned, has the advantage of being less singular than both the fermionic representation and the three-dimensional $s$-wave problem.

\subsection{Duality in effective field theory II:\\ Three-body sector}
I now tackle the problem of establishing duality relations between one-dimensional bosons and fermions in the three-body sector. In this case, interactions are already much weaker than in the two-body sector \cite{ValienteThreeBody}, and I will only consider the LO three-body force. However, these are especially important in two scenarios: (i) weakly repulsive (near non-interacting) Bose gases with negative, large scattering length, where the three-body interaction is dominant \cite{ValienteThreeBody,ValientePastukhov}; and (ii) attractive (single component) bosons with large, positive scattering length, whose binding energy scales as $\propto N^3$ (see Sect. \ref{Examples}) , with $N$ the particle number, if no three-body interaction is present and, as I will show, always act as repulsive, at low energies, and may stabilize quantum droplets. Higher multiparticle interactions, such as four-body processes, also feature interesting physical phenomena \cite{Nishida2010}, but shall not be considered here. 

Since the statistical interaction $W_{\mathrm{F}}$ does not generate three-body terms, all that is needed here is the unitary transformation between the bosonic three-body interaction $V_{\mathrm{B}}^{(3)}$ and its fermionic dual $V_{\mathrm{F}}^{(3)}=\mathcal{T}V_{\mathrm{B}}^{(3)}\mathcal{T}^{\dagger}$. In the three-body sector, the position representation of the STO is given by $S_3(\mathbf{x})=S(x_{12})S(x_{13})S(x_{23})$. I denote by $\ket{\chi_{\mathbf{k}}}$ the non-interacting fermionic three-body state with momenta $(k_1,k_2,k_3)$, i.e. $\ket{\chi_{\mathbf{k}}}=\mathcal{A}(\ket{\mathbf{k}})$, with $\mathcal{A}$ the antisymmetrization operator\footnote{\begin{align}
    \mathcal{A}{\ket{k_1,k_2,k_3}}\equiv (1/6)&[\ket{k_1,k_2,k_3}-\ket{k_1,k_3,k_2}+\ket{k_2,k_3,k_1}\nonumber\\
      &-\ket{k_2,k_1,k_3}+\ket{k_3,k_1,k_2}-\ket{k_3,k_2,k_1}].\nonumber \end{align}}. The momentum representation of the fermionic LO dual three-body interaction is given by
\begin{equation}
V_{\mathrm{F}}^{(3)}(\mathbf{k}',\mathbf{k})=\frac{g_0^{(3)}}{(2\pi)^6}\int \mathrm{d}\mathbf{q}' S_{\mathbf{q}',\mathbf{k}'}^{\mathrm{A}*}\int \mathrm{d}\mathbf{q} S_{\mathbf{q},\mathbf{k}}^{\mathrm{A}},\label{VF31}
\end{equation}
where total momentum conservation is assumed (there is an implicit factor of $2\pi\delta(K-K')$), together with a hard cutoff $\Lambda$, $g_0^{(3)}$ is the bosonic three-body coupling constant, $K=k_1+k_2+k_3$ is the total momentum, and where $S_{\mathbf{q},\mathbf{k}}^{\mathrm{A}}=\mathcal{A}(S_{\mathbf{q}-\mathbf{k}})$, antisymmetrized with respect to $\mathbf{k}$, with $S_{\mathbf{k}}$ the Fourier transform of the STO, i.e.
\begin{equation}
S_{\mathbf{k}}=\int \mathrm{d}\mathbf{x} S_3(\mathbf{x})e^{i\mathbf{k}\cdot \mathbf{x}}.
\end{equation}
The Fourier transform above can be performed analytically, giving, after tedious but straighforward algebra, 
\begin{equation}
S_{\mathbf{k}}=2\pi\delta(K)4\pi i\mathrm{P.V.}\left[\frac{\delta(Q_s)}{Q_t}-\frac{\delta(Q_t)}{Q_s}+\frac{\delta(Q_s+Q_t)}{Q_s}\right],\label{S3}
\end{equation}
where
\begin{align}
  Q_s&=\frac{1}{3}(2k_1-k_2-k_3)\\
  Q_t&=\frac{1}{3}(-k_1-k_2+2k_3).
\end{align}
Inserting Eq.~(\ref{S3}) into $S_{\mathbf{q},\mathbf{k}}^{\mathrm{A}}$ and integrating over $\mathbf{q}$, the following is obtained
\begin{equation}
\int \mathrm{d}\mathbf{q}S_{\mathbf{q},\mathbf{k}}^{\mathrm{A}}=i\pi^2\sum_{\mathrm{P}_{\ell m n}}(-1)^{\mathrm{P}}\dashint \mathrm{d}q\frac{1}{k_{mn}-q},\label{SA}
\end{equation}
where the sum above is over all permutations $(\ell m n)$ of $(123)$, and $k_{mn}=(k_m-k_n)/2$. I introduce Eq.~(\ref{SA}) into the three-body interaction, Eq.~(\ref{VF31}), and finally obtain
\begin{equation}
  V_{\mathrm{F}}^{(3)}(\mathbf{k}',\mathbf{k})=\frac{g_0^{(3)}}{(2\pi)^6}F^*\left(\frac{k_{12}'}{\Lambda},\frac{k_{23}'}{\Lambda}\right)F\left(\frac{k_{12}}{\Lambda},\frac{k_{23}}{\Lambda}\right)
\end{equation}
with
\begin{equation}
F(x,y)=2i\pi^2\ln\left|  \frac{(1+x)(1+y)(1-x-y)}{(1-x)(1-y)(1+x+y)}         \right|.
\end{equation}
Expanding the interaction to LO, the low energy dual interaction reads
\begin{equation}
  V_{\mathrm{F}}^{(3)}(\mathbf{k}',\mathbf{k})=\frac{g_0^{(3)}}{4\pi^2\Lambda^6}k_{12}'k_{13}'k_{23}'k_{12}k_{13}k_{23}+O(g_0^{(3)}\Lambda^{-8}),
\end{equation}
which yields, by comparing with Eq.~(\ref{Fermiong6}), a LO coupling constant
\begin{equation}
  g_6^{(3)}=\frac{g_0^{(3)}}{4\pi^2\Lambda^6}.
\end{equation}
Note that the non-trivial dependence of $g_0^{(3)}$ on the hyperspherical cutoff within the bosonic formalism can hinder practical computations with fermions as is. However, the cutoff dependence can be removed by instead realizing that the bosonic three-body interaction can be implemented with minimal subtraction \cite{ValienteThreeBody}. The three-body coupling constant can be replaced by its renormalized value, after choosing a subtraction point $-\mu^2$ \cite{Frederico}, $g_3^{(R)}=\pi\sqrt{3}\hbar^2\ln|\sqrt{\hbar^2/2m}Q_*/\mu|/m$ and subsequently define the LO bosonic three-body interaction as \cite{Frederico}
\begin{equation}
  V_{\mathrm{B}}^{(3)}=-g_3^{(R)}\left[\mu^2+z\right]G_0(-\mu^2),
\end{equation}
where total momentum conservation is implicitly assumed, $z=E+i\eta$ is the energy and $G_0(-\mu^2)$ is the three-body non-interacting Green's function at energy $-\mu^2$.

It is also possible to estimate the effects of the three-body interaction near the fermionization limit, that is, for low two-body scattering length $|a|$. The cutoff structure $\propto \Lambda^{-6}$ of the fermionic interaction suggests, from dimensional grounds, that its effect in perturbation theory should be of $O(a^6)$. For small scattering lengths the ground state is similar to that of the extended hard-rod model \cite{ValienteHard}, and agree to second order in $a$. Place a three-body system in a box of size $L$ with periodic boundary conditions. Using the exact wave function \cite{Girardeau}, and the renormalized coupling constant in a box \cite{ValienteThreeBody}, one obtains the correction to the three-body ground state energy due to the three-body force
\begin{equation}
  \langle V^{(3)} \rangle = \frac{32\pi^7}{\sqrt{3}}\frac{\hbar^2}{mL^2\ln\left|Q_*L/2\pi\right|}\left(\frac{a}{L}\right)^6+\ldots,\label{3bodycorrection}
\end{equation}
which is indeed of $O(a^6)$. This formally means that the three-body interaction near the free fermionic limit is very weak as compared to scattering length effects, of $O(a/L^3)$, and effective range ($r$) effects, of $O(a^2r/L^5)$. However, the numerical constant in Eq.~(\ref{3bodycorrection}) is very large, $32\pi^7/\sqrt{3}\approx 5\cdot 10^4$, of the order of $10^3$ times larger than the respective numerical constant for effective range effects, and is not necessarily negligible.

\subsection{Duality in effective field theory III:\\ Three-body bound states}
Here, I will use the EFT formalism for three-body bound states with LO two-body and three-body interactions for bosons and their fermionic dual. Since, for bound states, the fermionic problem is in the strong coupling (attractive) regime, it is computationally much harder to work with fermions and, therefore, the duality relations are very appealing for obtaining fermionic bound states via their bosonic representation. Three-body bound states in the momentum representation are most naturally studied within the formalism of Faddeev equations, which are derived for bosons and fermions in Appendix~\ref{AppFaddeev}.

\begin{figure}[t]
\begin{center}
\includegraphics[width=0.49\textwidth]{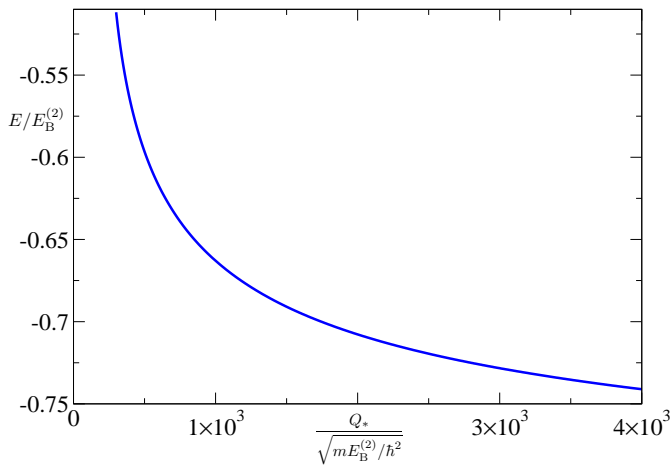}
\end{center}
\caption{Three-boson bound state energy $E$ with total momentum $K=0$ with LO two- and three-body interactions, in units of the three-boson binding energy with no three-body interaction $E_{\mathrm{B}}^{(2)}=mg_0^2/\hbar^2$, as a function of the three-body momentum scale $Q_*$. This state corresponds asymptotically ($Q_*\to\infty$) to the bound state of the Lieb-Liniger model, and not the deep state with asymptotic energy $-\hbar^2Q_*^2/2m$.}
\label{fig:ThreeBoson}
\end{figure}

\begin{figure}[t]
\begin{center}
\includegraphics[width=0.49\textwidth]{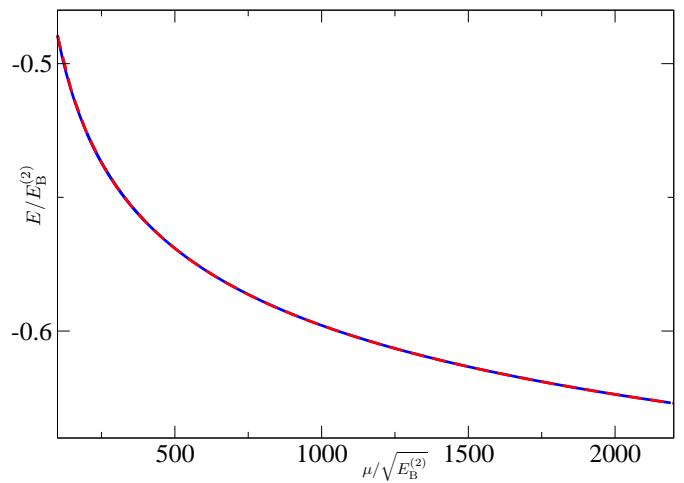}
\end{center}
\caption{Three-fermion bound state energy $E$ with total momentum $K=0$ with LO two-body interactions in the kernel subtraction scheme (blue solid line), as a function of the subtraction scale $\mu$ (subtraction energy $-\mu^2/2$), in units of the three-fermion binding energy with no three-body interaction $E_{\mathrm{B}}^{(2)}=mg_0^2/\hbar^2$. The red dashed line is a fit to the data (see text).}
\label{fig:ThreeFermion}
\end{figure}

The usefulness of the duality relations is patent when exploring bound state problems. For two particles, both bosonic and fermionic bound states are easy to calculate in the momentum representation. For three particles, bosonic bound states remain numerically tractable with either two-body forces only (which is exactly solvable in the position representation for both bosons and fermions) or two- and three-body forces. For the dual three-fermion bound states, however, even with only two-body forces in the Faddeev equations numerical convergence is rather slow, while including the three-body dual interaction in the bound state problem is even more problematic. To see this, I have solved the integral equation (\ref{TerMartirosyan}) for three bosons with fixed LO coupling constant $g_0$ and varying three-body momentum scale $Q_*$, and the corresponding fermionic dual, Eq.~(\ref{FermiTerMartirosyan}) with no three-body interaction ($Q_*\to \infty$). In the limit $\Lambda\to\infty$, the numerical solution of the fermionic bound state equation converges very slowly as the number of quadrature points is increased. Instead, I use the kernel subtraction method of Ref.~\cite{Frederico}, which is equivalent to a different kind of three-particle interaction, and subtraction point $-\mu^2/2$. The original problem is recovered in the limit $\mu^2\to\infty$. The results, fully converged for three bosons, and as a function of $\mu$ for fermions, are shown in Figs.~\ref{fig:ThreeBoson} and \ref{fig:ThreeFermion}, respectively. The convergence in the fermionic problem is logarithmically slow. Setting $x=\mu/\sqrt{E_{\mathrm{B}}^{(2)}}$, with $E_{\mathrm{B}}^{(2)}=mg_0^2/\hbar^2$ the exact three-body binding energy for $\mu^2\to\infty$, a logarithmic fit to the data of the form $E/E_{\mathrm{B}}^{(2)}(x)=a+b/\ln(dx)+c/\ln^2(dx)$ gives $a\approx -1.02$ (exact is $-1$). Convergence is so slow that, according to the fit, in order to achieve less than $10\%$ relative error in the binding energy, values of $\mu/\sqrt{E_{\mathrm{B}}^{(2)}}>10^{14}$ are required, for which numerical convergence is not achieved.

In short, the duality relations between bosons and fermions with two- and three-body interactions allow one to set the low-energy scattering parameters $a$ (scattering length) and $Q_*$ (three-body scale) for either bosons and fermions, and $N$-body bound states can be calculated using the bosonic representation, which is by far the simplest. If the target system is fermionic, the wave function $\ket{\chi}$ is obtained from the bosonic one $\ket{\psi}$ by using the STO.

\section{Dualities in multicomponent and spinful systems}\label{SectionDuality2}

\subsection{Statistical transmutation operators}
In systems with other internal degrees of freedom, be it different sublattices for continuum limits of tight-binding models, atomic levels, or spin, to name a few, STOs are not unique. This is easy to see in the two-body sector, where a bosonic two-body wave function $\psi(\mathbf{\xi}_1,\mathbf{\xi}_2)$ and a fermionic one $\chi(\mathbf{\xi}_1,\mathbf{\xi}_2)$ that are dual to one another via an STO $\mathcal{T}$ are related by
\begin{align}
  \psi(\mathbf{\xi}_1,\mathbf{\xi}_2)=\sum_{\mathbf{m}_1',\mathbf{m_2}'}&\bra{\mathbf{m}_1\mathbf{m}_2}\mathcal{T}(x_1-x_2)\ket{\mathbf{m}_1'\mathbf{m}_2'}\nonumber\\
  &\times\chi(x_1,x_2;\mathbf{m}_1',\mathbf{m}_2'),\label{chitopsimulti}
\end{align}
where I have used the locality of $\mathcal{T}$ and the fact that, since it is only concerned with particle exchange, it only depends on the relative coordinate $x_1-x_2$. Particle exchange in Eq.~(\ref{chitopsimulti}), together with unitarity, implies that the matrix elements $\mathcal{T}_{\mathbf{m}_1',\mathbf{m}_2'}^{\mathbf{m}_1,\mathbf{m}_2}(x_{12})$, of $\mathcal{T}$
\begin{equation}
  \mathcal{T}_{\mathbf{m}_1',\mathbf{m}_2'}^{\mathbf{m}_1,\mathbf{m}_2}(x_{12})=\bra{\mathbf{m}_1\mathbf{m}_2}\mathcal{T}(x_1-x_2)\ket{\mathbf{m}_1'\mathbf{m}_2'},
\end{equation}
can only have two different spatial dependences, either a constant or proportional to the signum distribution $S(x_1-x_2)$. For $\mathbf{m}_1=\mathbf{m}_2$ and $\mathbf{m}_1'=\mathbf{m}_2'$, clearly the only possible spatial dependence is proportional to $S(x_1-x_2)$, since particle exchange only affects the exchange of spatial coordinates. In order to show the non-uniqueness of STOs in the multichannel case, it suffices to consider a general two-component system, with single-particle components $\ket{1}$ and $\ket{2}$. Using the following matrix ordering for two particles, $\{\ket{11},\ket{12},\ket{21},\ket{22}\}$, it is easy to see that the following local operators $\mathcal{T}_1$ and $\mathcal{T}_2$ are valid STOs, $\mathcal{T}_1(x_1-x_2)=S(x_1-x_2)\mathbb{I}$, with $\mathbb{I}$ the $4\times 4$ identity operator and $\mathcal{T}_2(x_1-x_2)=\mathcal{A}S(x_1-x_2)/4$, with

\begin{equation}
  \mathcal{A}=
\begin{pmatrix}
  1 & i & i & 1 \\
  i & 1 & -1 & -i \\
  i & -1 & 1 & -i \\
  1 & -i & -i & 1
\end{pmatrix}
,
\end{equation}
which are related via a simple (symmetric and regular) unitary transformation, which adds no new physical content.
Note, again, that the STOs are valid for any two-component system regardless of the kinetic energy and interaction details of the Hamiltonian. As I will show in the next subsection, $\mathcal{T}_1$, which allows for a straightforward generalization to many particles, see Eq.~(\ref{diagonalSTO}), is important from the physical point of view regarding duality relations, and will be used in two relevant two-component models below.

\subsection{Duality in the continuum limit of the Su-Schrieffer-Heeger model}
The Su-Schrieffer-Heeger (SSH) model \cite{SSHpaper1,SSHpaper2} is a paradigmatic tight-binding model of condensed matter physics in one spatial dimension with a plethora of interesting emergent features. Besides its topological nature, which has been even shown experimentally using trapped ultracold atoms \cite{BlochSSH} admitting a zero-energy in-gap edge state for a perfect boundary \cite{Gadway2016,Rice1982,Duncan2018}, it provides a Dirac-like dispersion in its continuum limit near half filling for spinless fermions. Bose-Fermi duality in the hard-core limit holds in the SSH model on the lattice, since tunneling is nearest-neighbour only. However, its continuum limit has a different structure and duality will take a form very different from the na\"ive continuum limit. The non-interacting SSH model has the lattice Hamiltonian
\begin{equation}
  H_0=\sum_jt_j(c_{j+1}^{\dagger}c_j+\mathrm{H.c.}),\label{SSHHamiltonian}
\end{equation}
where $t_j=t+(-1)^j\delta/2$, and $c_j$ ($c_j^{\dagger}$) the annihilation (creation) operator of either a spinless boson or fermion at site $j$. Fermions with Hamiltonian (\ref{SSHHamiltonian}) and bosons with the same Hamiltonian together with the interaction term
\begin{equation}
  V= \frac{U}{2}\sum_jn_j(n_j-1), \hspace{0.1cm} U\to +\infty,\label{SSHInt}
\end{equation}
where $n_j=c_j^{\dagger}c_j$, are dual to each other. In the continuum limit near quasi-momentum $kd=\pi/2$, with $d$ the lattice spacing, the first quantized version of the single-particle Hamiltonian (\ref{SSHHamiltonian}) takes the form
\begin{equation}
  H_0=
\begin{pmatrix}
 0 & -i\hbar v\partial_x-i\delta\\
 -i\hbar v \partial_x+i\delta & 0  \\
\end{pmatrix}
,\label{SSHcontinuum}
\end{equation}
where the velocity $v=-2td/\hbar$, and each component corresponds to a sublattice ($1$ and $2$) of the SSH model. For fermions, this is well defined provided they are non-interacting or weakly interacting. Bosons, on the other hand, must be strongly interacting in order for their excitations, which are not single particle entities, to behave according to Hamiltonian (\ref{SSHcontinuum}). To extract the continuum Hamiltonian describing the original, microscopic bosons, Bose-Fermi duality in the continuum is required. I use the diagonal STO, Eq.~(\ref{diagonalSTO}), which gives the following statistical interaction in the position representation
\begin{equation}
  W(\mathbf{x})=-2i\hbar v\sum_{i<j}S(x_{ij})\delta(x_{ij})\mathcal{M}_{ij},\label{StatisticalSSH}
\end{equation}
where $\mathcal{M}_{ij}$ acts on the sublattice degrees of freedom of particles $i$ and $j$, given by
\begin{equation}
  \mathcal{M}_{ij}=\ket{11}(\bra{21}-\bra{12})+\ket{22}(\bra{12}-\bra{21})+\mathrm{H.c.},\label{Mij}
\end{equation}
where $\ket{n_1n_2}\equiv \ket{n_1}_i\otimes\ket{n_2}_j$. Observe that Hermiticity of $W$ in Eq.~(\ref{StatisticalSSH}) is guaranteed by the properties of Shirokov's algebra, in particular the anticommutativity of signum and delta distributions, Eq.~(\ref{anticommutator}). The position representation of $W$, Eq.~(\ref{StatisticalSSH}), is not very useful unless the exact solution is used, as usual. Its momentum representation using a cutoff requires, just as in the usual non-relativistic cases, some microscopic knowledge, see Appendix~\ref{Appendix}. To see this, take two particles and write $W(\mathbf{k}',\mathbf{k})=w(\mathbf{k}',\mathbf{k})\mathcal{M}_{ij}$ as
\begin{align}
w(k',k)&=-2i\hbar v \int \frac{\mathrm{d}q}{2\pi}S_{q-k'}\nonumber\\
&=2i\hbar v \int \frac{\mathrm{d}q}{2\pi}S_{k-q}.
\end{align}
Without a cutoff in the above integrals, there is no problem {\it a priori}, since they both vanish. Introducing a cutoff $\tilde{\Lambda}$ and expanding the results, however, gives
\begin{equation}
w(\mathbf{k}',\mathbf{k})=-\frac{4\hbar v}{\pi}\frac{k'}{\tilde{\Lambda}}=-\frac{4\hbar v}{\pi} \frac{k}{\tilde{\Lambda}}.\label{expandW}
\end{equation}
The cutoff $\tilde{\Lambda}$ is not the same as the cutoff ($\Lambda$) one introduces when solving, say, the two-body problem. All that Eq.~(\ref{expandW}) says is that the coupling constant is negative and proportional to $1/\Lambda$. This is because the above expression, Eq.~(\ref{expandW}), is an expansion of already regular, separable interactions. The fact that the expressions in Eq.~(\ref{expandW}) are not equal to each other before $\tilde{\Lambda}\to\infty$ is not a problem, since they only need to agree in that limit. To fix the coupling constant in Eq.~(\ref{expandW}) with the appropriate cutoff $\Lambda$, it is simplest to consider the massless limit ($\delta=0$), and use the fact that, since the fermionic dual problem to the bosonic problem with either statistical interaction in Eq.~(\ref{expandW}) is renormalizable (in fact, free), so is the bosonic problem.
The two-body scattering problem is solved in detail in Appendix~\ref{AppendixSSH}, where I show that the statistical interaction as a function of the actual cutoff $\Lambda$ in the two-boson problem, that renormalizes the theory and puts fermions and bosons in one-to-one correspondence, Eq.~(\ref{expandW}), takes the form $w(k',k)=-4\pi\hbar v k'/\Lambda$ or $w(k',k)=-4\pi\hbar v k/\Lambda$. The symmetric choice, $w(k',k)=-(4\pi\hbar v/\Lambda)(k'+k)$, also gives identical results.

The introduction of finite interactions in the bosonic problem is of course possible and it is more complicated than in the fermionic dual (see Appendix~\ref{AppendixSSH}), which makes the duality relations very useful in this case. It is easiest to introduce the simplest interactions for fermions, which are of the form
\begin{equation}
  V(x_{12})=g_0\delta(x_{12})\hat{O},\label{VO}
\end{equation}
with $\hat{O}=\ket{12}\bra{12}+\ket{21}\bra{21}$. The dual bosonic interaction $\tilde{V}$ is given, in the momentum representation, by
\begin{equation}
  \tilde{V}(k',k)=\frac{\pi^2mg_0}{4\hbar^2\Lambda^2}k'k\hat{O},\label{VBO}
\end{equation}
and therefore the full boson-boson interaction, with the choice of statistical interaction $\propto k$, is given by $\tilde{V}-4\pi\hbar v (k/\Lambda)\mathcal{M}_{ij}$.

\subsection{Duality between non-relativistic spin-$1/2$ fermions and two-component bosons}
I consider now non-relativistic spin-$1/2$ fermions with Dirac delta, even-wave interactions, corresponding to Yang's model \cite{Yang1967}. Its Hamiltonian is given by
\begin{equation}
  H=\sum_{i=1}^N\frac{p_i^2}{2m}+g_0\sum_{i<j=1}^N\delta(x_i-x_j).\label{Yang}
\end{equation}
The dual bosonic theory to the above model was obtained using pseudopotentials in Ref.~\cite{Girardeau2004}. I show here its EFT construction, and how to discretize it, for numerical purposes, on a grid (lattice), and subsequently present results for three particles that clearly show duality in the continuum limit.

I now discuss the bosonic dual to Yang's model in Eq.~(\ref{Yang}). In the triplet (spin-symmetric states) collision channels, bosonic statistics implies symmetric spatial exchange. The only term in the dual bosonic Hamiltonian in this channel is an even-wave two-body statistical interaction. From Eq.~(\ref{Wgen}), this is seen to be a linearly divergent constant as function of the cutoff. Therefore, interactions in the triplet ($\ket{\tau_1}=\ket{\uparrow\uparrow}$, $\ket{\tau_2}=\ket{\downarrow\downarrow}$, $\ket{\tau_3}\equiv (1/\sqrt{2})(\ket{\uparrow\downarrow}+\ket{\downarrow\uparrow})$) channels are given by
\begin{equation}
  \bra{k'\tau_i'}W_{\mathrm{B}}\ket{k\tau_i}=\gamma\delta_{i,i'}, \hspace{0.1cm} \gamma\to +\infty.
\end{equation}
Above, the Kronecker delta indicates that the statistical interaction due to the STO (\ref{diagonalSTO}) induces no spin flips. The statistical interaction between a triplet and the singlet channel also vanishes. In the singlet channel (spin antisymmetric $\ket{s}=(1/\sqrt{2})(\ket{\uparrow\downarrow}-\ket{\downarrow\uparrow})$), bosonic statistics dictates spatial exchange antisymmetry. Therefore, the singlet statistical interaction is odd-wave, and Eq.~(\ref{Wgen}) gives $\bra{k's}W_{\mathrm{B}}\ket{ks}=-(\pi\hbar^2/m\Lambda)k'k$. The structure of the STO (\ref{diagonalSTO}) generates no three- and higher-body terms in the statistical interaction. The interaction in Yang's model (\ref{Yang}) only affects singlet states, a fact that carries over to the bosonic dual. The (odd-wave) bosonic dual interaction in the singlet channel is obtained by matching, as in the spinless case, and gives, as $\Lambda\to\infty$,
\begin{equation}
  \bra{k'\alpha'}V_{\mathrm{B}}\ket{k\alpha}=\frac{\pi^2g_0}{4\Lambda^2}\delta_{\alpha,s}\delta_{\alpha,\alpha'}.
\end{equation}

\begin{figure}[t]
\begin{center}
\includegraphics[width=0.49\textwidth]{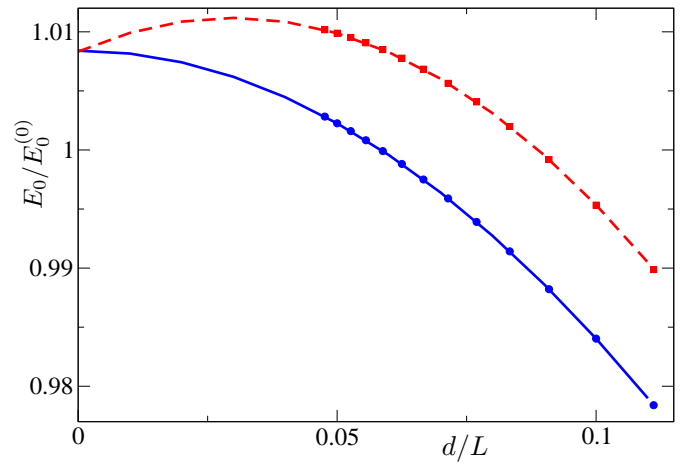}
\end{center}
\caption{Ground state energy $E_0$ in units of the continuum limit ($d\to 0$) of the fermionic non-interacting ground state energy $E_0^{(0)}$ of three fermions (bosons) with spins $\uparrow\uparrow\downarrow$ (components $1$,$1$,$2$) corresponding to Yang's model and its dual, in a box of size $L$ as a function of the lattice spacing for $mLg_0/\hbar^2=1/10$. Blue circles and red squares correspond, respectively, to fermions and bosons. Solid blue (red dashed) line is a fit of Eq.~(\ref{fitfermion}) (Eq.~(\ref{fitboson})) to the fermionic (bosonic) data.}
\label{fig:Yang}
\end{figure}

A straightforward discretization of Yang's model (fermions), with lattice spacing $d$, is given by the Hubbard Hamiltonian
\begin{equation}
  H_d=-J\sum_{j,\sigma=\uparrow,\downarrow}(c_{j+1\sigma}^{\dagger}c_{j\sigma}+\mathrm{H.c.})+2JN+U\sum_{j}n_{j\uparrow}n_{j\downarrow},
\end{equation}
where $c_{j\sigma}$ ($c_{j\sigma}^{\dagger}$) annihilates (creates) a fermion with spin $\sigma$ at site $j$, $n_{j\sigma}=c_{j\sigma}^{\dagger}c_{j\sigma}$ is the number operator at site $j$, and $N=\sum_{j\sigma}n_{j\sigma}$ is the total number operator. The continuum limit is attained for $J=\hbar^2/2md^2$ and $U=g_0/d$, with $d\to 0$. For bosons, one has two components $b_{j1}$ and $b_{j2}$, corresponding to spin-$\uparrow$ and spin-$\downarrow$, respectively, in this dual representation. Since the triplet interactions are hardcore, intracomponent interactions are hardcore. Since, for the remaining triplet state, the interaction is also hardcore, the bosonic Hamiltonian contains hardcore even-wave interactions $V_{\mathrm{HC}}$ of the form 
\begin{equation}
  V_{\mathrm{HC}}=U_{\infty}\left[\frac{1}{2}\sum_{j\sigma}n_{j\sigma}(n_{j\sigma}-1)+\sum_jn_{j1}n_{j2}\right], \hspace{0.1cm} U_{\infty}\to +\infty,
\end{equation}
with the sums over $\sigma=1,2$. The singlet states interact via an odd-wave interaction. In the first quantization, and denoting the relative site index by $j_r=j_1-j_2$, this interaction $V_o$ is given by
\begin{equation}
  \bra{j_r'}V_o\ket{j_r}=\frac{v}{2}(\delta_{j_r,1}+\delta_{j_r,-1})(\delta_{j_r,j_r'}-\delta_{j_r,-j_{r}'}),
\end{equation}
with \cite{ValienteZinnerEFT}
\begin{equation}
  v=-\frac{2J}{1-d/a},
\end{equation}
where $a$ is the scattering length, related to $g_0$ as $g_0=-2\hbar^2/ma$. In the second quantization, this is realized by defining an exchange operator $\hat{O}^{\mathrm{ex}}$ as
\begin{equation}
  \hat{O}^{\mathrm{ex}}=\sum_{j_1,j_2}b_{j_12}^{\dagger}b_{j_21}^{\dagger}b_{j_11}b_{j_22},
\end{equation}
and the odd-wave singlet-singlet $V_{s}$ interaction takes the form
\begin{equation}
  V_{s}=\frac{v}{2}\sum_j\left(n_{j,1}n_{j+1,2}+n_{j,2}n_{j+1,1}\right)\left(1-\hat{O}^{\mathrm{ex}}\right).
\end{equation}
I obtain the ground state of three fermions (bosons), two of them with spin-$\uparrow$ (component $1$) and one with spin-$\downarrow$ (component $2$) on a lattice with varying spacing $d$ and open boundary conditions. The length of the continuum target system is $L=(L_s+1)d$, where $L_s$ is the number of lattice sites used in the calculation. I fix the value of $L$ and vary $d$ according to $d=L/(L_s+1)$. For fermions, the best fit for the ground state energy $E_0^{\mathrm{F}}(d)$ is quadratic in $d$,
\begin{equation}
  E_0^{\mathrm{F}}(d)=E_0+ad^2+bd^4,\label{fitfermion}
\end{equation}
while for bosons, which contain multiple strong interactions, the finite-$d$ scaling shows non-monotonic behaviour and a fit of this form to the ground state energy $E_0^{\mathrm{B}}(d)$ works well
\begin{equation}
  E_0^{\mathrm{B}}(d)=\tilde{E}_0+\tilde{a}d+\tilde{b}d^2+\tilde{c}d^3.\label{fitboson}
\end{equation}
In Fig.~\ref{fig:Yang} I show the values of the ground state energy $E_0$ in units of the non-interacting ground state energy $E_0^{(0)}$ for three interacting fermions with spins $\uparrow\uparrow\downarrow$ in a box of size $L$ with varying lattice spacing, together with fits (\ref{fitfermion}) and (\ref{fitboson}) to the numerical data. Clearly, the results for fermions and dual bosons are identical in the continuum limit within extrapolation errors. 

\section{Duality as a gauge interaction. Extension to anyons}\label{SectionGauge}
The STOs presented here are all unitary and the emergent interactions in either the bosonic or fermionic representation do not couple the center of mass and relative coordinates, preserving Galilean invariance when this is present in the original system. Therefore, it is possible to find a gauge interaction which is antisymmetric upon particle exchange \cite{Jackiw} which, upon elimination, is equivalent to the introduction of the STO. I also consider one-dimensional anyons, which were incorrectly introduced as a gauge theory decades ago \cite{Rabello1996}, as pointed out by Aglietti and co-workers in Ref.~\cite{Jackiw}. I will show that, although anyons cannot be introduced by means of a conventional gauge interaction, a particle label-dependent term can do the job. More importantly, I derive the statistical interaction for the anyonic STO, obtain its EFT description and show that it is renormalizable. I will focus on the spinless one-component case for simplicity and concreteness.

I denote by $a_j(\mathbf{x})$ the gauge interaction for particle $j$, and $\Phi$ the function satisfying
\begin{equation}
 -i\hbar\partial_{x_j}\Phi(\mathbf{x})-a_j(\mathbf{x})\Phi(\mathbf{x})=0, \hspace{0.1cm} j=1,2,\ldots,N.\label{ajEq}
\end{equation}
Since the STO is simply $\mathcal{T}(\mathbf{x})=S_N(\mathbf{x})$, one has $\Phi(\mathbf{x})=S_N(\mathbf{x})$. Introducing this into Eq.~(\ref{ajEq}) and multiplying the resulting equation by $S_N(\mathbf{x})$ on the right, together with the properties (\ref{anticommutator}) and (\ref{Ssquare}) of Shirokov's algebra, one obtains
\begin{equation}
  a_j(\mathbf{x})=2i\hbar\sum_{\ell(\ell\ne j)}S(x_{j\ell})\delta(x_{j\ell}).
\end{equation}

I consider now one-dimensional anyons with statistical angle $\phi$ \cite{Keilmann2010}. The spatial part of the local STO $\mathcal{T}_{\phi}(\mathbf{x})$ is given by
\begin{align}
  \mathcal{T}_{\phi}(\mathbf{x})&=i\exp\left[-i\phi\sum_{j<\ell}S(x_{j\ell})\right]\nonumber\\
  &=i\prod_{j<\ell}\left[\cos\phi-iS(x_{j\ell})\sin\phi\right].\label{STOanyons}
\end{align}
A gauge interaction $a_j^{\phi}$ of the following form can be defined,
\begin{equation}
a_j^{\phi}(\mathbf{x})=-\hbar\sum_{\ell=1}^{N}\left[\sin(2\phi_{j\ell})+2i\sin^2\phi S(x_{j\ell})\right]\delta(x_{j\ell}),\label{aphi}
\end{equation}
where the sum above is restricted to $\ell\ne j$ and $\phi_{j\ell}=\phi$ for $j<\ell$ and $\phi_{j\ell}=-1$ for $j>\ell$. Note that, in Eq.~(\ref{aphi}), it was necessary to include a term $\propto \sin(2\phi_{j\ell})$ which depends on labelling of the particles, in order to preserve the antisymmetry. Therefore, I would be reluctant to consider Eq.~(\ref{aphi}) a proper gauge interaction for indistinguishable particles. Of course, the anyonic STO (\ref{STOanyons}) induces a statistical interaction $W_{\phi}=[\mathcal{T}_{\phi},H_0]\mathcal{T}_{\phi}^{\dagger}$. For non-relativistic particles, this includes two- ($W_{\phi}^{(2)}$) and three-body parts ($W_{\phi}^{(3)}$). Since I will only consider the two-body problem below, I only write down explicitly the two-body part, given by
\begin{align}
  W_{\phi}^{(2)}(\mathbf{x})&=-i\frac{\hbar^2}{m}\sin(2\phi)\sum_{i<j}\left[\delta'(x_{ij})+2\delta(x_{ij})\partial_{x_{ij}}\right]\nonumber\\
  &+\sin^2\phi W(\mathbf{x}),\label{Wphipos}
\end{align}
where $W=W_{\pi/2}$ is the statistical interaction of the Bose-Fermi mapping, Eq.~(\ref{WF1channel}). The three-body statistical interaction $W_{\phi}^{(3)}$ is obtained analogously. As in the case of Bose-Fermi duality, Eq.~(\ref{Wphipos}) is highly formal due to the distributional nature of the statistical interaction. A momentum-space regularization is desirable, and it is given by
\begin{equation}
  W_{\phi}^{(2)}(k',k)=\frac{\hbar^2}{m}\sin(2\phi)(k+k')+\sin^2\phi \left[g_0^{\infty}-\frac{\pi\hbar^2}{m\Lambda}k'k\right],\label{Wphimom}
\end{equation}
where $g_0^{\infty}=4\hbar^2\Lambda/m\pi$.

Consider now two-body scattering with the statistical interaction $W_{\phi}^{(2)}$ of Eq.~(\ref{Wphimom}). The $T$-matrix can be split as
\begin{equation}
  T(z;k',k)=\tau_0(z)+\tau_+(z)k+\tau_{-}(z)k'+\tau_{+-}(z)k'k,
\end{equation}
yielding two coupled systems of algebraic equations, namely
\begin{align}
  \tau_0&=g_0+g_0I_0\tau_0+g_{01}I_2\tau_{-},\\
  \tau_{-}&=g_{01}+g_{01}I_0\tau_0+g_{11}I_2\tau_{-},
\end{align}
and
\begin{align}
  \tau_{+}&=g_{01}+g_0I_0\tau_++g_{01}I_2\tau_{+-},\\
  \tau_{+-}&=g_{11}+g_{01}I_0\tau_{+}+g_{11}I_2\tau_{+-},
\end{align}
where I have defined $g_{0}=g_0^{\infty}\sin^2\phi$, $g_{01}=\hbar^2\sin(2\phi)/m$, $g_{11}=-\sin^2\phi\pi\hbar^2/m\Lambda$ and
\begin{equation}
  I_n=I_n(z)=\int \frac{\mathrm{d}q}{2\pi}\frac{q^n}{z-\hbar^2q^2/m}.
\end{equation}
As $\Lambda\to\infty$, setting $z=\hbar^2k^2/m+i\eta$ gives, for the first system,
\begin{align}
  \tau_0&=\frac{-i\hbar^2|k|\cos^2\phi\left[4\sin^2\phi+\sin^2(2\phi)\right]/2m}{\cos^2\phi+\left[4\sin^2\phi+\sin^2(2\phi)\right]/2},\label{tau0}\\
  \tau_{-}&=\frac{2\hbar^2\tan\phi}{m}\left[1-\frac{im}{2\hbar^2|k|}\tau_0\right],\label{tauminus}
\end{align}
which affects the scattering of identical bosons. A similar solution is obtained for $\tau_{+}$ and $\tau_{+-}$, which affect the scattering of identical fermions. The on-shell $T$-matrix (for bosons) is given by
\begin{equation}
  T^{\mathrm{on}}=\left[e^{-i\phi}\tau_0+\frac{2\hbar^2|k|}{m}\sin\phi\right]\sec \phi.
\end{equation}
It is interesting to note from Eqs.~(\ref{tau0},\ref{tauminus}) that, for low statistical angle $\phi$, the lowest order source of scattering for bosons is the term $g_{01}k'$, which gives a contribution to the $T$-matrix that is linear in $\phi$, which can be included perturbatively to lowest order. Also, as one would expect, anyonic statistics itself does not generate any bound states, since the $T$-matrix has no poles.

\section{Conclusions}\label{SectionConclusions}
In this article, an extended version of an accompanying Letter \cite{Letter}, I have presented a detailed account of the most general duality relations between bosons and fermions in one spatial dimension. These are valid for arbitrary low-energy interactions, including multiparticle forces among more than two bodies, spin or multicomponent structure and single-particle Hamiltonian or dispersion in the continuum. For spinless non-relativistic systems, it has been shown that the low-energy physics of interacting bosons and fermions are equivalent to one another, a fact that would be difficult to prove using a scattering theory approach, especially for more than two particles. The results have been extended to systems with arbitrary internal structure or spin, and regularized in a manner that is computationally tractable, for few bodies in the momentum representation and for many particles using a lattice discretization -- amenable to density matrix renormalization group \cite{Schollwoeck2005} calculations -- or a real space regularization -- amenable to quantum Monte Carlo methods -- for both Galilean systems and otherwise. One-dimensional anyons can be treated in a completely analogous way and have been briefly discussed.

These results have a number of direct consequences. Firstly, any system of bosons (fermions) in one dimension, in continuous space with low-energy interactions, regardless of the particular details of the single-particle Hamiltonian and internal structure, can be treated in either the original or dual representations. This allows to choose the most convenient particle statistics, depending on how weakly or strongly coupled they are. For example, spinless bosons with effectively attractive two-body interactions and repulsive three-body interactions can form one-dimensional quantum droplets \cite{ValienteOhberg}, and therefore the same is true for spinless fermions (whose low-energy interactions have been recently manipulated by means of $p$-wave Feshbach resonances \cite{Hulet2020}) with their dual Hamiltonian. Quantum droplets of two-component bosons \cite{Petrov2015}, which have been created and observed in three dimensions \cite{Tarruell2018,Fattori2018}, are also predicted to occur in one dimension \cite{Petrov2016,Astrakharchik2018,Morera2020,Morera2020b}. The conditions for the formation of spinful fermionic droplets, which would be in the strong-coupling limit, can be directly inferred from the bosonic multicomponent theory \cite{Petrov2016} and the duality relations. Bose-Fermi duality also applies to coupled wires, whether genuinely continuous or as continuum limits of tight-binding quantum ladders \cite{Cazalilla2011,Celi2014,Fallani2016,Kolkowitz2017}, where each wire represents a different component (spin or true components can also be included), and these are coupled already at the single-particle level -- a fact that only affects the form and nature of the statistical interaction. With optical lattice-based ladders \cite{Fallani2016,Kolkowitz2017}, it is also important to note that, while the general Bose-Fermi mapping does not apply on a tight-binding model, it does apply to the more accurate continuous description with a non-relativistic single-particle dispersion and an external periodic potential. I leave other potentially interesting consequences and applications of the results obtained here to the ingenuity of our colleagues.

\appendix

\section{Two-body coupling constants to next-to-leading order}\label{AppendixNLO}
Here I present the solution to the spinless two-boson and two-fermion problems with LO and NLO interactions. For two bosons, the interaction is given by
\begin{equation}
  V_{\mathrm{e}}(k',k)=g_0+g_2(k'^2+k^2).\label{VbosonApp}
\end{equation}
Since I am dealing with identical particles, I will obtain the scattering phase shifts via the reaction operator ($R$-matrix), which is simpler as it does not involve imaginary quantities. The reaction matrix uses standing-wave boundary conditions, as opposed to the $T$-matrix, which uses incident-scattering wave boundary conditions. For identical particles, these are equivalent since incident wave functions are standing waves ($\cos(kx)$ for bosons and $\sin(kx)$ for fermions). The Lippmann-Schwinger equation for the $R$-matrix reads
\begin{equation}
  R(z)=V+VG_0^{\mathrm{sw}}(E)R(z),
\end{equation}
where $G_0^{\mathrm{sw}}(E)=[G_0(E+i0^+)+G_0(E+i0^-)]/2\equiv \mathrm{Re}(G_0(E+i0^+))$.

The $R$-matrix for non-relativistic bosons with interaction (\ref{VbosonApp}) can be written as
\begin{equation}
  R_{\mathrm{e}}(z;k',k)=R_0(z)+R_{20}(z)k'^2+R_{02}(z)k^2+R_{22}(z)k'^2k^2.
\end{equation}
The $R$-matrix above can be calculated analytically, and after tedious but straightforward algebra, one observes that $R_{20}$, $R_{02}$ and $R_{22}$ all vanish as the cutoff $\Lambda\to\infty$, while $R_0(z)$, with $z=\hbar^2k^2/m$, takes the renormalized form
\begin{equation}
  R_0(z)=\frac{2\hbar^2}{m}\left(-\frac{1}{a}+\frac{1}{2}rk^2\right),\label{Rbosons}
\end{equation}
where $a$ and $r$ are, respectively, the scattering length and effective range. Defining the constant $g_{\mathrm{R}}=-2/a$, renormalization is attained by fixing the bare coupling constants $g_0$ and $g_2$ in Eq.~(\ref{VbosonApp}) as
\begin{align}
  \frac{mg_2(\Lambda)}{\hbar^2}&=-\frac{\pi}{\Lambda}\left[1+\sqrt{-\frac{g_{\mathrm{R}}+\pi\Lambda}{r\Lambda^2}}\right]^{-1},\label{g2App}\\
  \frac{mg_0(\Lambda)}{\hbar^2}&=\left(\frac{\Lambda}{3\pi}+\frac{1}{r}\right)\left[\Lambda \frac{mg_2(\Lambda)}{\hbar^2}\right]^2+\Lambda\pi\left(1+\frac{\Lambda mg_2(\Lambda)}{\pi\hbar^2}\right)^2.\label{g0App}
\end{align}

The fermion interaction is given by
\begin{equation}
  V_{\mathrm{o}}(k',k)+W_{\mathrm{F}}(k',k)=g_1k'k+g_3k'(k'^2+k^2)k,\label{VfermionApp}
\end{equation}
where the odd-wave statistical interaction is already included in $g_1$. The $R$-matrix for two fermions can be written as
\begin{align}
  R_{\mathrm{o}}(z;k',k)&=k'k\left[R_1(z)+R_{31}(z)k'^2+R_{13}(z)k^2\right.\nonumber\\
    &\left.+R_{33}(z)k'^2k^2\right].
\end{align}
This can be calculated analytically as well and, analogously to the bosonic case, one has vanishing $R_{31}$, $R_{13}$ and $R_{33}$ in the limit $\Lambda\to\infty$, while $R_1(z)$, with $z=\hbar^2k^2/m$, takes the renormalized form
\begin{equation}
  R_1(z)=-\frac{2\hbar^2}{m}\left(-\frac{1}{a}+\frac{1}{2}rk^2\right)^{-1}.\label{Rfermions}
\end{equation}
Renormalization is attained by fixing the fermionic coupling constants $g_1$ and $g_3$ as
\begin{align}
  \frac{mg_3(\Lambda)}{\hbar^2}&=-\frac{3\pi}{\Lambda^3}\left[1-\sqrt{\left(\frac{\pi g_{\mathrm{R}}}{4\Lambda}+1\right)\frac{3}{1-\pi r \Lambda/4}}\right],\label{g3App}\\
    \frac{mg_1(\Lambda)}{\hbar^2}&=-\frac{1}{\Lambda}\frac{3\pi}{1-\pi r\Lambda/4}+\frac{\left[\Lambda^3mg_3(\Lambda)/\hbar^2\right]^2}{5\pi\Lambda}.\label{g1App}
\end{align}
Note that for both the bosonic and fermionic problems, the effective range is restricted to negative values, as observed in Eqs.~(\ref{g2App},\ref{g3App}). This, while a strong constraint, is not unphysical. In the recent experiment of Ref.~\cite{Hulet2020} with quasi-one-dimensional polarized fermions with odd-wave interactions, the measured effective range is negative, and therefore the corresponding low-energy physics can be described by zero-range EFT.

\begin{figure}[t]
\begin{center}
\includegraphics[width=0.49\textwidth]{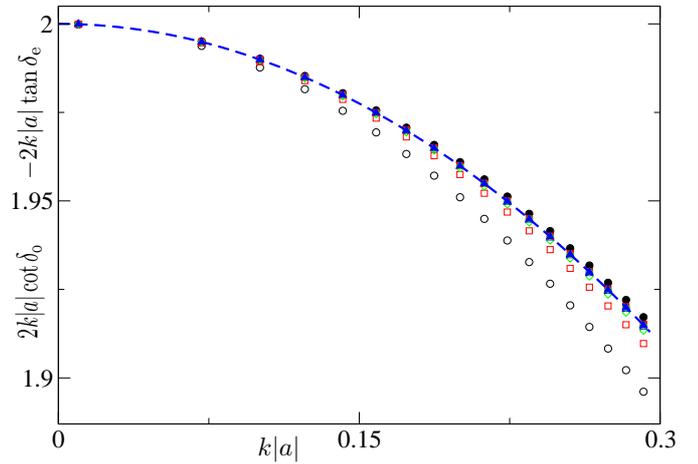}
\end{center}
\caption{Scattering phase shifts to LO+NLO for spinless bosons and fermions as functions of the relative momentum with $r<0$ and $r/a=1$. Open (filled) symbols correspond to spinless bosons (fermions). From bottom to top (top to bottom), $\Lambda|a|=10^2,10^3,10^4$ and $10^5$. Blue dashed line is the exact result in the limit $\Lambda\to\infty$.}
\label{fig:PhaseShifts}
\end{figure}

The bosonic and fermionic theories are dual in virtue of Eqs.~(\ref{Rbosons},\ref{Rfermions}). To see this, I first note that the $R$-matrix for bosons (fermions) is independent of momentum (only depends on momentum proportionally to $k'k$), just as in the case of LO scattering, which means that the scattering states of bosons (fermions) take the form $\psi(x)=\cos(k|x|+\delta_{\mathrm{e}})$ ($\chi(x)=S(x)\sin(k|x|+\delta_{\mathrm{o}})$). The scattering phase shifts $\delta_{\mathrm{e}}$ for the bosonic $R$-matrix take on values
\begin{equation}
  -k\tan\delta_{\mathrm{e}}=\left[-\frac{1}{a}+\frac{1}{2}rk^2\right],\label{psboson}
\end{equation}
while the fermionic phase shifts are given by
\begin{equation}
  k\cot\delta_{\mathrm{o}}=\left[-\frac{1}{a}+\frac{1}{2}rk^2\right].\label{psfermion}
\end{equation}
For duality to hold, the phase shifts must be related as $\delta_{\mathrm{o}}=\delta_{\mathrm{e}}+\pi/2$, which is satisfied by Eqs.~(\ref{psboson},\ref{psfermion}). In Fig.~\ref{fig:PhaseShifts} , I plot the corresponding phase shifts for bosons and fermions with different values of the cutoff, using the full reaction matrices calculated with the bare coupling constants in Eqs.~(\ref{g2App},\ref{g0App},\ref{g3App},\ref{g1App}), together with the exact expression in the limit $\Lambda\to\infty$ (Eqs.~(\ref{psboson},\ref{psfermion})). There, it is observed that, as the cutoff is increased, both bosonic and fermionic phase shifts converge to the desired value.

\section{Pseudopotentials and regularization}\label{Appendix}
In this appendix, I consider some technical details about the duality relations, via the statistical transmutation operator, between non-interacting bosons and strongly-interacting fermions.

Throughout the manuscript, position-represented interactions are fully regularized and renormalized by regarding distributions and their products as members of Shirokov's algebra, introduced in Sect.~\ref{SectionShirokov}. As is usual with position-represented exact pseudopotentials, these are not always convenient unless the exact solution is known \cite{Olshanii2001}, and different, equivalent representations can yield wildly different results when approximate solutions are sought \cite{Olshanii2001}. A straightforward example of this fact can be constructed by considering $N$ spin-$\uparrow$ and one spin-$\downarrow$ fermions in three spatial dimensions interacting via a LO $s$-wave potential in the unitary limit, i.e. with infinite scattering length -- the Fermi polaron problem \cite{Chevy2010}. An ansatz containing up to one particle-hole term \cite{Chevy2006} yields, variationally, infinite energy if the Fermi-Huang-Yang pseudopotential is used \cite{Huang1987,Tan2005,Valiente2012}. If one uses instead non-perturbative cutoff regularization and renormalization, the variational ground state energy lies very close to Monte Carlo results \cite{Chevy2006,Prokofev2008}. These two ``different'' approaches correspond to two particular choices in the possible family of equivalent, exact pseudopotentials of Ref.~\cite{Olshanii2001}. Note also that regularized-renormalized representations of pseudopotentials, including the Fermi-Huang-Yang interaction \cite{Huang1987}, are not necessarily Hermitian, as they aim to reproduce the correct right eigenstates of an interacting system. What is important is that they lead to unitary evolution which, in cutoff regularization schemes, is guaranteed in the large cutoff limit where Hermiticity is typically restored.

Cutoff regularization and renormalization in the momentum representation has the advantage that it can be used both perturbatively and non-perturbatively in a consistent manner. While transformations using momentum cutoffs are well defined and yield exact results, expansions of the resulting expressions in powers of the (inverse) cutoff need to be dealt with carefully, and either minimal microscopic input, as is the case below for only one coupling constant, or a more involved approach otherwise (see Appendix~\ref{AppendixNLO}), is required. To illustrate this, the momentum representation of the non-relativistic statistical interaction $W$ can be obtained from its position representation (\ref{WF1channel}) as
\begin{equation}
  \bra{k'}W\ket{k}=\frac{2\hbar^2}{m\pi}\dashint\mathrm{d}q\frac{k+q}{q-k'}.
\end{equation}
Introducing a cutoff $\tilde{\Lambda}$ to regularize the integral above, one obtains
\begin{equation}
  \bra{k'}W\ket{k}=\frac{4\hbar^2\tilde{\Lambda}}{m\pi}+\frac{2\hbar^2}{m\pi}(k+k')\ln\left|\frac{\tilde{\Lambda}-k'}{\tilde{\Lambda}+k'}\right|.\label{Wapp1}
\end{equation}
So far, the statistical interaction (\ref{Wapp1}) is exact in the limit $\tilde{\Lambda}\to\infty$, that is, one can solve the problem at hand for finite $\tilde{\Lambda}$ and the results are exact as $\tilde{\Lambda}$ is taken to infinity. The even- and odd-wave projections of $W$ are given by
\begin{align}
   W_{\mathrm{e}}(k',k)&=\frac{4\hbar^2\tilde{\Lambda}}{m\pi}+\frac{2\hbar^2}{m\pi}k'\ln\left|\frac{\tilde{\Lambda}-k'}{\tilde{\Lambda}+k'}\right|,\label{Weapp}\\
  W_{\mathrm{o}}(k',k)&=\frac{2\hbar^2}{m\pi}k\ln\left|\frac{\tilde{\Lambda}-k'}{\tilde{\Lambda}+k'} \right|.\label{Woapp}
 \end{align}
For large $\tilde{\Lambda}$, the even-wave statistical interaction (\ref{Weapp}) is just the point hard-core interaction and the momentum dependent term can be dropped. Even though this might sound trivial, it actually is not: it is possible to only retain the constant term $\propto \tilde{\Lambda}$, unmodified, because that term alone leads to a well-defined, renormalized $T$-matrix. The odd-wave statistical interaction (\ref{Woapp}) can be expanded as
\begin{equation}
  W_{\mathrm{o}}(k',k)=-\frac{4\hbar^2}{m\pi\tilde{\Lambda}}kk'+ O(\tilde{\Lambda}^{-3}).\label{Woexp}
\end{equation}
Now consider the Lippmann-Schwinger equation for the $T$-matrix for the expanded interaction
\begin{equation}
  T(z;k',k)=-\frac{4\hbar^2}{m\pi\tilde{\Lambda}}kk'-\frac{4\hbar^2}{m\pi\tilde{\Lambda}}k'\int \frac{\mathrm{d}q}{2\pi} \frac{q}{z-\hbar^2q^2/m}T(z;q,k),
\end{equation}
which is solved by $T(z;k',k)=\tau(z)k'k$, yielding
\begin{equation}
  \tau(z)=-\left[\frac{m\pi\tilde{\Lambda}}{4\hbar^2}+\int \frac{\mathrm{d}q}{2\pi}\frac{q^2}{z-\hbar^2q^2/m}\right]^{-1}.\label{tauz}
\end{equation}
Since the free boson problem is well defined, and the odd-wave statistical interaction maps, by construction, fermions to non-interacting bosons, the $T$-matrix must be renormalizable. Using a cutoff ($\Lambda$) regularization of the integral in Eq.~(\ref{tauz}) and imposing renormalizability, together with the non-existence of further momentum scales, one immediately has
\begin{equation}
\tilde{\Lambda}=\frac{4\Lambda}{\pi^2},
\end{equation}
which implies that the cutoff regularization of the odd-wave part of the statistical interaction is given by
\begin{equation}
  W_{\mathrm{o}}(k',k)=-\frac{\pi\hbar^2}{m\Lambda}kk'.
\end{equation}
Note that the difference between the cutoffs $\tilde{\Lambda}$ and $\Lambda$ is natural. The interaction in Eq.~(\ref{Woapp}) is regular, and no cutoff is needed as $\tilde{\Lambda}$ acts, for fermions, as a momentum scale that needs to be sent to infinity at the end of the calculation. When expanding the interaction, Eq.~(\ref{Woexp}), the resulting low-energy potential is singular, and yields UV divergences in the calculation of the $T$-matrix. This means that the expansion must be supplemented by regularization, that is, the expansion actually reads
\begin{equation}
  W_{\mathrm{o}}(k',k)=-\frac{4\hbar^2}{m\pi\tilde{\Lambda}}kk'\theta(|k|-\Lambda),
\end{equation}
with $\Lambda$ a new momentum scale that by no means has to be equal to $\tilde{\Lambda}$. However, no new momentum scales can be introduced in the problem, since no new physics is introduced, which means that $\tilde{\Lambda}$ and $\Lambda$, while different, must be linearly related.

\section{Faddeev equations}\label{AppFaddeev}
In this Appendix, I provide details of the derivation of the three-body Faddeev equations for non-relativistic bosonic and fermionic bound states.

The three-body $T$-matrix $T$ is split into four components using the Faddeev decomposition as \cite{Bazak}
\begin{equation}
  T=T^1+T^2+T^3+U,
\end{equation}
where $T^i$ ($i\ne j\ne \ell \ne i$) and $U$ satisfy
\begin{align}
  T^i(z)&=t^i(z)+t^i(z)G_0(z)(T^j(z)+T^{\ell}(z))\nonumber \\
        &+t^i(z)G_0(z)U(z),\label{Faddeev1}\\
  U(z)&=u(z)+u(z)G_0(z)(T^i(z)+T^j(z)+T^{\ell}(z)).\label{Faddeev2}
\end{align}
Above, $G_0(z)$ is the three-body non-interacting Green's function, $t^i(z)$ is a two-body spectator $T$-matrix, i.e.
\begin{equation}
t^i(z)=V_{j\ell}+V_{j\ell}G_0(z)t^i(z),
\end{equation}
and $u(z)$ is the three-body $T$-matrix in the absence of two-body interactions, i.e.
\begin{equation}
  u(z)=V^{(3)}+V^{(3)}G_0(z)u(z).
\end{equation}
At a bound state energy, the (operator-valued) residues of the Faddeev components are denoted by $M^i(z)$ and $R(z)$ for $T^i(z)$ and $U(z)$, respectively, and equations (\ref{Faddeev1}) and (\ref{Faddeev2}) for $M^i$ and $R$ remain identical except for the inhomogeneous terms which now disappear. Introducing Eq.~(\ref{Faddeev2}) into Eq.~(\ref{Faddeev1}), one easily obtains
\begin{align}
M^i(z)&=t^i(z)G_0(z)(M^j(z)+M^{\ell}(z))\nonumber \\
        &+t^i(z)G_0(z)u(z)(M^i(z)+M^j(z)+M^{\ell}(z)).\label{FaddeevBound}
\end{align}  

I now particularize Eq.~(\ref{FaddeevBound}) to bosons with LO two- and three-body interactions. Without loss of generality, I will assume that the total momentum vanishes ($K=0$) within the arguments of functions. The spectator two-body $T$-matrix is given, for $z=E<0$, by
\begin{align}
  \bra{\mathbf{k}'}t^i(z)\ket{\mathbf{k}}&=(2\pi)^2\delta(k_i-k_i')\delta(K-K')t_2\left(E-\frac{3\hbar^2k_i^2}{4m}\right),\\
  t_2(E)&=\frac{1}{\frac{1}{g_0}+\pi\sqrt{\frac{m}{\hbar^2|E|}}}.
\end{align}
The three-body $T$-matrix in the absence of two-body interactions $u(z)$ (for $z=E<0$) is given by \cite{ValienteThreeBody}
\begin{equation}
\bra{\mathbf{k}'}u(z)\ket{\mathbf{k}}=2\pi\delta(K-K')\frac{2\pi\sqrt{3}\hbar^2}{m}\frac{1}{\ln\left|\frac{E_*}{E}\right|},\label{uz1}
\end{equation}
where $E_*\equiv -\hbar^2Q_*^2/2m$ is the location of the three-body Landau pole, with $Q_*$ becoming the only momentum scale in the problem, which defines the strength of the three-body interaction. The matrix elements of the Faddeev components $M^i(z)$ can be simplified considerably due to the momentum-independent nature of $t_2(E)$ and $u(E)$. By working out the momentum representation of Eq.~(\ref{FaddeevBound}), it is observed that
\begin{equation}
\bra{\mathbf{k}'}M^i(z)\ket{\mathbf{k}}=\tilde{M}^i(z,k_i'),
\end{equation}
i.e. only a function of the spectator particle's momentum $k_i'$. Using bosonic symmetry and defining $\tilde{M}^i(z,k_i')\equiv \tilde{M}(k)$, where I have also redefined the dummy momentum variable $k_i'=k$ and obviated the energy dependence, Eq.~(\ref{FaddeevBound}) takes the simple form
\begin{align}
  &\tilde{M}(k)=-t_2\left(E-\frac{3}{4}\frac{\hbar^2k^2}{m}\right)\times\nonumber\\
  &\int \frac{\mathrm{d}q}{\pi}\left[\frac{1}{|E|+\epsilon(k,q,k+q)}-\frac{3}{2}u(E)I(E,k)I(E,q)\right]\tilde{M}(q),\label{TerMartirosyan}
\end{align}
where $u(z)$ is given by Eq.~(\ref{uz1}) after removing the factor $2\pi\delta(K-K')$, 
\begin{equation}
I(E,k)=\int \frac{\mathrm{d}q}{2\pi}\frac{1}{E-\epsilon(k,q,k+q)},
\end{equation}
and where I have defined $\epsilon(k_1,k_2,k_3)=\sum_{i=1}^3\hbar^2k_i^2/2m$.

I now derive the bound state equation for the fermionic case. The Faddeev components $M^i(z)$ have the following momentum representation, after removing the factor $2\pi \delta(K-K')$,
\begin{equation}
M^i=F^i(k_i)k_{j\ell},
\end{equation}
where $i\ne j\ne \ell\ne i$, and $k_{j\ell}=(k_j-k_{\ell})/2$. Fermionic statistics dictates that
\begin{align}
  F^i(k_j)&=-F^j(k_i),\\
  F^i(k_{\ell})&=F^{\ell}(k_i),\\
  F^j(k_{\ell})&=-F^{\ell}(k_j),
\end{align}
which, after defining $F^i(k_i)\equiv F(k)$, simplifies the Faddeev equations to
\begin{equation}
F(k)=\alpha\left(E-\frac{3}{4}\frac{\hbar^2k^2}{m}\right)\int \frac{\mathrm{d}q_j}{2\pi}\mathcal{K}(k,q_j)F(q_j),\label{FermiTerMartirosyan}
\end{equation}
where $\alpha(E)$ is given by
\begin{equation}
  \alpha(E)=-\frac{1}{m^2g_0/4\hbar^2+\sqrt{m|E|/\hbar^2}/2},
\end{equation}
and 
\begin{align}
  \mathcal{K}(k,q_j)&=\frac{q_{j\ell}(q_{ij}-q_{i\ell})}{E-\epsilon(k,q_j,k+q_j)}\nonumber\\
  &+\Gamma_{\Lambda}(E)\tilde{I}_{\Lambda}(E,k)\Omega_{\Lambda}(E,q_j).
\end{align}
Above, I have defined
\begin{align}
  \Omega_{\Lambda}(E,q_j)&=\int \frac{\mathrm{d}q_i}{2\pi}\frac{q_{ij}q_{i\ell}q_{j\ell}(q_{j\ell}+q_{ij}-q_{i\ell})}{E-\epsilon(q_i,q_j,q_i+q_j)},\\
  \tilde{I}_{\Lambda}(E,k)&=\frac{1}{4}\int \frac{\mathrm{d}q_{j\ell}}{2\pi}\frac{q_{j\ell}^2(9k^2-4q_{j\ell})}{E-\epsilon(k,q_j,k+q_j)}.\\
\frac{1}{\Gamma_{\Lambda}(E)}&=\frac{1}{g_6^{(3)}}+\int \frac{\mathrm{d}q_i}{2\pi}\int \frac{\mathrm{d}q_j}{2\pi}\frac{(q_{ij}q_{i\ell}q_{j\ell})^2}{|E|+\epsilon(\mathbf{q})}.
\end{align}

\section{Scattering in the continuum limit of the SSH model}\label{AppendixSSH}
In this Appendix, I present the renormalization of the two-boson problem dual to fermions in the continuum limit of the SSH model, and its statistical interaction.

 For simplicity and without loss of generality, I consider the massless case ($\delta=0$ in Eq.~(\ref{SSHcontinuum})). Right- (R) and left-moving (L) single particle states with energy $\hbar v k$ are given by
\begin{align}
  \ket{k\mathrm{R}}&=\frac{1}{\sqrt{2}}\left[\ket{1}+\ket{2}\right]\ket{k},\\
  \ket{-k\mathrm{L}}&=\frac{1}{\sqrt{2}}\left[\ket{1}-\ket{2}\right]\ket{-k}.
\end{align}

The matrix structure ($\mathcal{M}_{ij}$) of the statistical interaction (\ref{StatisticalSSH},\ref{Mij}) only gives non-zero matrix elements for $\bra{\mathrm{L}\mathrm{R} k_1'k_2'}W\ket{\mathrm{L}\mathrm{R} k_1k_2}$ and $\bra{\mathrm{R}\mathrm{L} k_1'k_2'}W\ket{\mathrm{R}\mathrm{L} k_1k_2}$. For the $\propto k$ choice in Eq.~(\ref{expandW}), the $T$-matrix, after separating the factor indicating total momentum conservation, takes the simple form $T_{ij}(z;k',k)=\tau_{\mathrm{a};i,j}(z)k$, with $z=2\hbar v \bar{k}+i\eta$. Denoting $w(k',k)=w_ak/2$, the Lippmann-Schwinger equations read
\begin{align}
  \tau_{\mathrm{a};\mathrm{LR}}(z)&=-w_a-w_a\tau_{\mathrm{a};\mathrm{LR}}(z)\int_{-\Lambda}^{\Lambda}\frac{\mathrm{d}q}{2\pi}\frac{q}{2\hbar v(\bar{k}+q)+i\eta}\label{tauLR}\\
  \tau_{\mathrm{a};\mathrm{RL}}(z)&=w_a+w_a\tau_{\mathrm{a};\mathrm{RL}}(z)\int_{-\Lambda}^{\Lambda}\frac{\mathrm{d}q}{2\pi}\frac{q}{2\hbar v(\bar{k}-q)+i\eta}.\label{tauRL}
\end{align}
The two equations above, with $w_a\propto -1/\Lambda$, are renormalizable if $w_a=-2\pi\hbar v/\Lambda$, i.e. the coupling constant in Eq.~(\ref{expandW}) is given by $-4\pi\hbar v/\Lambda$. The solutions to Eqs.~(\ref{tauLR}) and (\ref{tauRL}) are given by $\tau_{\mathrm{a};\mathrm{LR}}(z)=4i\hbar v/\bar{k}$ and $\tau_{\mathrm{a};\mathrm{RL}}(z)=-4i\hbar v/\bar{k}$ which give, on-shell, $T_{\mathrm{RL}}^{\mathrm{on}}=T_{\mathrm{LR}}^{\mathrm{on}}=-4i\hbar v$. The symmetric choice of statistical interaction $w(k',k)=-(4\pi\hbar v/\Lambda)(k'+k)$ gives identical results, albeit the solution of its (now channel-coupled) Lippmann-Schwinger equation is more involved. The two-boson scattering state $\ket{\psi_{k}}$ (for a fixed center of mass momentum $K$) is constructed as 
\begin{equation}
  \ket{\psi_{k}}=\left[1+G_0(z)T(z)\right]\ket{\psi_{k}^{(0)}},\label{psik}
\end{equation}
where $\ket{\psi_{k}^{(0)}}=(1/\sqrt{2})(\ket{k\mathrm{RL}}+\ket{-k\mathrm{LR}})$ is the bosonic incident state. It is straightforward to see that the position representation of (\ref{psik}) is simply $\ket{\psi_k(x)}=S(x)\ket{\chi_k^{(0)}(x)}$, where $\ket{\chi_k^{(0)}}=(1/\sqrt{2})(\ket{k\mathrm{RL}}-\ket{-k\mathrm{LR}})$ is its free fermionic dual. Choosing the other version ($\propto k'$), or the symmetric one ($\propto k'+k$), of the statistical interaction (\ref{expandW}) gives identical results.

Note that, because of the first derivatives in the Hamiltonian, bosons in the $\ket{\mathrm{RR}}$ and $\ket{\mathrm{LL}}$ channels do not interact. This implies that free bosonic states composed of only right- or left- moving particles are eigenstates of the Hamiltonian. The fact is that, since the statistical interaction does not couple these states, the bosonic states of the form $\mathcal{T}\ket{\chi}$ are also eigenstates and, therefore, nothing is {\it a priori} wrong. However, this means that further input -- the bosonic wave functions should be the dual ones -- is required (notice that the same issues occur for free fermions). A neat way to solve this problem is to introduce infinitesimaly weak contact interactions in the $\ket{\mathrm{RR}}$ and $\ket{\mathrm{LL}}$ channels. The interaction commutes with the Hamiltonian, and their common eigenstates are just $\mathcal{T}\ket{\chi}$ \cite{ValienteFlat}.

I introduce now fermion-fermion interactions of the form of Eq.~(\ref{VO}). In the momentum representation, the interaction reads $\bra{k'i'j'}V\ket{kij}=g_0\bra{i'j'}\hat{O}\ket{ij}$, with
\begin{align}
  \bra{\mathrm{LR}}\hat{O}\ket{\mathrm{LR}}&=\bra{\mathrm{RL}}\hat{O}\ket{\mathrm{RL}}=\frac{1}{2},\\
  \bra{\mathrm{LR}}\hat{O}\ket{\mathrm{RL}}&=\bra{\mathrm{RL}}\hat{O}\ket{\mathrm{LR}}=-\frac{1}{2}.
\end{align}
The interaction (\ref{VO}) is renormalizable with finite $g_0$. All elements of the $T$-matrix are constant (momentum independent) and, while coupled, their Lippmann-Schwinger equations are straightforward. Defining $\bra{i'j'}T\ket{ij}=T_{ij}^{i'j'}$, one gets two independent systems of coupled algebraic equations. For concreteness, one of these reads
\begin{align}
  T_{\mathrm{LR}}^{\mathrm{LR}}&=\frac{g_0}{2}+\frac{g_0}{2}I_{\mathrm{LR}}T_{\mathrm{LR}}^{\mathrm{LR}}-\frac{g_0}{2}I_{\mathrm{RL}}T_{\mathrm{LR}}^{\mathrm{RL}},\\
T_{\mathrm{LR}}^{\mathrm{RL}}&=\frac{g_0}{2}-\frac{g_0}{2}I_{\mathrm{LR}}T_{\mathrm{LR}}^{\mathrm{RL}}+\frac{g_0}{2}I_{\mathrm{RL}}T_{\mathrm{LR}}^{\mathrm{LR}}.
\end{align}
Above, $I_{ij}$ are defined as (with $z=2\hbar\bar{k} v+i\eta$)
\begin{align}
  I_{\mathrm{LR}}&=\int \frac{\mathrm{d}q}{2\pi}\frac{1}{2\hbar v(\bar{k}+q)+i\eta}=-\frac{i}{4\hbar v},\\
  I_{\mathrm{RL}}&=\int \frac{\mathrm{d}q}{2\pi}\frac{1}{2\hbar v(\bar{k}-q)+i\eta}=-\frac{i}{4\hbar v}.
\end{align}
The other system of two coupled equations is completely analogous. For bosons, on the other hand, the dual interaction $\tilde{V}$ is, in the momentum representation, given by Eq.~(\ref{VBO}). Together with the statistical interaction, the system of equations obtained from the Lippmann-Schwinger equation is rather formidable, and its renormalization, which is ensured by the duality transformation, is quite tedious. For instance, with the choice of statistical interaction $\propto k$, one has
\begin{align}
  T_{\mathrm{LR}}^{\mathrm{LR}}&=\tau_+k+\tau_{+-}k'k,\\
  T_{\mathrm{LR}}^{\mathrm{RL}}&=\tilde{\tau}_{+-}k'k.
\end{align}
It can be shown that $\tau_{+-}\to 0$ when $\Lambda\to\infty$ as $\sim \Lambda^{-1}$. This UV behaviour, however, must be kept in order to produce a finite $T$-matrix before taking the limit $\Lambda\to\infty$. It then follows that $\tau_{+}$ is finite and takes the functional form
\begin{equation}
  \tau_{+}=-\frac{1}{\bar{k}/4i\hbar v +\alpha},
\end{equation}
where $\alpha$ is real and momentum independent.

\end{document}